\documentclass[Afour,sagev,times]{sagej}
\usepackage{moreverb,url}

\usepackage[colorlinks,bookmarksopen,bookmarksnumbered,citecolor=red,urlcolor=red]{hyperref}

\usepackage{latexsym,ifthen,rotating,calc,textcase,booktabs,color,endnotes}
\usepackage{amsfonts,amssymb,amsbsy,amsmath,amsthm}
\usepackage{bm}
\usepackage[errorshow]{tracefnt}
\usepackage{color,xcolor}
\usepackage{soul}
\soulregister\cite7
\soulregister\ref7
\soulregister\eqref7

\usepackage{bookmark}

\newcommand\BibTeX{{\rmfamily B\kern-.05em \textsc{i\kern-.025em b}\kern-.08em
T\kern-.1667em\lower.7ex\hbox{E}\kern-.125emX}}

\def\volumeyear{2021}

\begin{document}

\runninghead{Liu, Song, Xu, Zhang and Xie}

\title{Discrete Boltzmann Modeling of Plasma Shock Wave}

\author{Zhipeng Liu\affilnum{1,2}, Jiahui Song\affilnum{1,3}, Aiguo Xu\affilnum{1,4,5}, Yudong Zhang\affilnum{6} and Kan Xie\affilnum{3}}

\affiliation{\affilnum{1}Laboratory of Computational Physics, Institute of Applied Physics and Computational Mathematics, Beijing 100088, China\\
\affilnum{2}Department of Physics, School of Science, Tianjin Chengjian University, Tianjin 300384, China\\
\affilnum{3}School of Aerospace Engineering, Beijing Institute of Technology, Beijing 100081, China \\
\affilnum{4}HEDPS,Center for Applied Physics and Technology, and College of Engineering, Peking University, Beijing 100871, China\\
\affilnum{5}State Key Laboratory of Explosion Science and Technology, Beijing Institute of Technology, Beijing 100081, China\\
\affilnum{6}School of Mechanics and Safety Engineering, Zhengzhou University, Zhengzhou 450001, China
}

\corrauth{Aiguo Xu, Kan Xie}

\email{Xu\_Aiguo@iapcm.ac.cn, xiekan@bit.edu.cn}

\begin{abstract}
Plasma shock waves widely exist and play an important role in high-energy-density environment, especially in the inertial confinement fusion. Due to the large gradient of macroscopic physical quantities and the coupled thermal, electrical, magnetic and optical phenomena, there exist not only hydrodynamic non-equilibrium (HNE) effects, but also strong thermodynamic non-equilibrium (TNE) effects around the wavefront. In this work, a two-dimensional single-fluid discrete Boltzmann model is proposed to investigate the physical structure of ion shock. The electron is assumed inertialess and always in thermodynamic equilibrium. The Rankine-Hugoniot relations for single fluid theory of plasma shock wave is derived. It is found that the physical structure of shock wave in plasma is significantly different from that in normal fluid and somewhat similar to that of detonation wave from the sense that a peak appears in the front. The non-equilibrium effects around the shock front become stronger with increasing Mach number. The charge of electricity deviates oppositely from neutrality in upstream and downstream of the shock wave. The large inertia of the ions causes them to lag behind, so the wave front charge is negative and the wave rear charge is positive. The variations of HNE and TNE with Mach number are numerically investigated. The characteristics of TNE can be used to distinguish plasma shock wave from detonation wave.
\end{abstract}


\maketitle

\section{Introduction}
Shock waves widely exist in nature and engineering fields, such as in supersonic flows and detonation. Plasma shock wave, which propagates in plasma, is a kind of shock wave accompanied by electromagnetic effects. One of the important application of the plasma shock wave is laser driven initial confinement fusion (ICF), in which materials are ionized by high-energy laser and strong shock waves are formed to compress pellets and heat fuel\cite{craxton2015direct, betti2016inertial, zhou2017rayleigh1, zhou2017rayleigh2, wang2021bell, xue2020explosion}.

From a macro perspective, shock waves are generally regarded as strong discontinuities formed by the superposition of a series of weak disturbances. But from the micro and mesoscopic perspectives, the shock wave has a finite width and possesses fine physical structures\cite{Pham1989Science,Xu2015APS,Liu2017SCPMA,Xu-Chapter2,Xu2021AAS,Xu2021AAAS}. The physical structure and propagation mechanism of shock wave in the macroscopic sense in neutral fluids have been studied for decades and well understood. However, with the development of measuring technology, the kinetic structure of shock wave and its variation with key parameters such as Mach number ($\rm{Ma}$) and species mass ratio between different particles, etc., are attracting more attention \cite{Liu2016FOP,Liu2017PRE,Yan2013FOP,Lin2014CTP,Lin2014PRE,Lin2016CNF,Zhang2016CNF,Gan2018PRE,Xu2018FOP,Zhang2018FOP,Zhang2019CPC,Lin2021PRE,Qiu2020PoF}. Especially, due to its complexity, the kinetic behavior of shock wave in plasma is still kept far from clear.

In view of the complexity of plasma shock wave, the one-dimensional collisional plasma shock wave was studied in most literatures. The understanding of one-dimensional collisional plasma shock waves is of fundamental reference value for understanding more complex plasma shock waves. In literatures, most of the work were focused on a shock wave propagating in a fully ionized plasma with no external forces. Both hydrodynamic and kinetic models were used to study the steady-state structure and relevant macroscopic quantities around shock front. In earlier studies, two-fluid plasma model based on Navier-Stokes (NS) equations was adopted by several authors and some interesting physical images were obtained. For example, Jukes\cite{jukes1957structure} investigated the velocity and temperature distribution through the shock wave, and found the electron temperature change more gradually than ion through a larger distance. Meanwhile, the ion temperature rises to a maximum that slightly higher than the final equilibrium temperature. However, in that study the electric field is neglected,
which means those  results are valid only when the scale of flow behavior under investigation is much larger than the Debye length (characteristic scale of charge separation) $\lambda_{d}$. At the same time, the electron viscosity and ion thermal conductivity were neglected. Jaffrin and Probstein\cite{Jaffrin1964Structure} gave the typical structure of plasma shock front by combining two-fluid NS equations with Poisson equation, and the transport coefficients given by Spitzer\cite{Spitzer1956book} and Braginskii\cite{braginskii1965reviews} were adopted. The result shows an ion shock wave imbedded in wider electron thermal layer when $\rm{Ma} > 1.12$. The thickness of ion shock is proportional to downstream ion mean free path, and there exists a preheating layer in front of the ion shock where electron temperature is higher than ion. Moreover, when the shock is strong ($\rm{Ma} = 10$), a precursor electric shock layer appears upstream, in which electric effects interact with flow. Ramirez\cite{ramirez1993nonlocal} studied the nonlocal electron heat relaxation effects by generalizing the results of Jaffrin with arbitrary ionization number $Z$. The results prove that nonlocal electron heat transport widens the preheating layer and smoothes the electron temperature profile. Hu\cite{yemin2003properties} considered the plasma electric characters such as current, field and charge around the shock front. It was found that a weak current carried by the shock could obviously affect the shock strength. Masser\cite{masser2011shock} developed semi-analytic solutions by using a two-temperature model, and found the boundary between continues and discontinues solutions depended on the upstream Mach number. In the literatures above, only one type of ion has been considered. Simakov\cite{simakov2014electron,simakov2016hydrodynamicI,simakov2016hydrodynamicII} then generalized the Braginskii electron fluid description to multi-ion plasma shock waves.

Though much effort and progress have been made by using hydrodynamic method, such as the NS equations. It should be noted that this treatment is mainly based on continuous medium hypothesis, and is valid only when the Knudsen number (defined as mean free path divide by a characteristic length scale) $\rm{Kn} \ll 1$. In other words, hydrodynamic method is valid only when considering behavior in a scale large enough. It is understandable that such a treatment will be challenged when considering behavior in small scale, for example, in scale for showing shock structure, especially strong shock structure with extremely high physical gradient. A more specific case  is that the structure of imbedded ion shock should be explored. Moreover, the $\rm{Kn}$ of shock wave become larger as $\rm{Ma}$ increases.
The Knudsen number can also be regarded as the relative thermodynamic relaxation time with respect to the time scale of flow under consideration. Since the time scale for shocking is very small, there exist extremely strong non-equilibrium effects near shock front. For the convenience of description, the non-equilibrium described by hydrodynamic theory is referred to as Hydrodynamic Non-Equilibrium (HNE), and the non-equilibrium described by kinetic theory due to deviating from thermodynamic equilibrium is referred to as Thermodynamic Non-Equilibrium (TNE). It is clear that the HNE is only one small part of TNE\cite{zhang2020two}. Additionally, the coefficients of viscosity and heat conduction (the two parameter describing the TNE) in NS are usually determined by experience or experiment.

As a more fundamental description method, kinetic theory based on distribution function is more suitable for investigating the fine structure of plasma shock. The most widely used kinetic model is the Boltzmann equation, in which the collision between particles is described by the Boltzmann operator\cite{cercignani1988boltzmann}, as shown in Eq. \eqref{Eq:Boltzmann Operator}
\begin{equation}\label{Eq:Boltzmann Operator}
Q(f,f) = \int_{ - \infty }^{ + \infty } {\int_0^{4\pi } {({f^ * }f_1^ *  - f{f_1})} } g\sigma d\bf{\Omega} d\bf{{v_1}}
\end{equation}
Among them, $f = f\left( {{\bf{r}},{\bf{v}},t} \right)$, ${f_1} = f\left( {{\bf{r}},{{\bf{v}}_1},t} \right)$, ${f^*} = f\left( {{\bf{r}},{{\bf{v}}^*},t} \right)$, and $f_1^* = f\left( {{\bf{r}},{\bf{v}}_1^*,t} \right)$ are the distribution function of the particle velocity ${\bf{v,}}{{\bf{v}}_{\bf{1}}}{\bf{,}}{{\bf{v}}^*}{\bf{,v}}_{\bf{1}}^*$ at position $\bf{r}$ at time $t$ respectively. There are extremely complex high-dimensional integral terms in the Boltzmann collision operator which generally cannot be solved analytically except in special cases\cite{krook1977exact}. For the plasma system, the particles interact through the Coulomb potential. At this case, the Boltzmann operator diverges\cite{villani2002review}, and the radius of action of the Coulomb potential needs to be truncated, and approximate collision operators such as Landau or Fokker-Planck collision operator based on small-angle scattering are introduced\cite{landau1958kinetic, rosenbluth1957fokker}. However, the collision operator after the introduction of approximation is still extremely complicated. In order to avoid the high computational cost caused by direct calculation, simplified models such as the Bhatnagar-Gross-Krook (BGK) operator\cite{bhatnagar1954model} are often used for further simplification.

Tidman\cite{Tidman1958Structure} employed Boltzmann-Fokker-Planck equations by assuming that the distribution function of ion is a bi-Maxwellian form. This form was put forward by Mott-Smith\cite{Mott-Smith1951PR}. The idea is that the distribution function transitions from the upstream equilibrium state to the downstream equilibrium state, which allows the distribution function deviate from equilibrium state around the shock wave front. However, Tidman's\cite{Tidman1958Structure} work neglect the electron thermal conductivity, which plays an important role in forming the electron preheating layer because of the small electron-ion mass ratio. Greenberg and Treve\cite{greenberg1960shock} first investigated the self-induced electric field caused by charge separation by using Mott-Smith bi-Maxwellian form, but only considered the momentum exchange between ions and electrons. Such a treatment is insufficient because dissipative effects such as ion/electron viscosity and thermal conductivity play important roles in maintaining the continuity structure of shock front. Casanova\cite{casanova1991kinetic} computed the ionic Fokker-Planck equation, combining with the electron heat equation. The result shows that ionic viscosity and thermal conduction are much larger than classic transport coefficients assumed in hydrodynamic simulation, and the effective shock width is comparable to the width of the electronic preheating layer, which means a hundredfold increase over the classic value. Videl\cite{vidal1993ion} then used the same model and discovered that the broaden of shock front is because of the energetic ions streaming from the hot and dense plasma into the cold. Keenan and Simakov\cite{keenan2017deciphering} adopted a high-fidelity Vlasov-Fokker-Planck code to investigate the shock structure with Mach number and different ion composition. By comparing the result with multi-ion hydrodynamic simulation, they find that the kinetic width of shock saturate for $\rm{Ma} \gg 1$, and the asymptotic value depends on the upstream lighter species concentration.

It is easy to find that the study on plasma shock wave is facing a dilemma. From one side, the hydrodynamic model based on NS is insufficient for describing the kinetic behavior of plasma. From the other side, the full kinetic model is too complicated to solve for most cases. It is meaningful to develop some coarse-grained kinetic model whose physical description capability is in between the hydrodynamic and fully kinetic models. The Discrete Boltzmann Modeling (DBM) method is one of methods for constructing such coarse-grained kinetic models\cite{Xu2015APS,Xu-Chapter2}. It can be regarded as a variant hybrid of the Lattice Boltzmann Method (LBM)\cite{succi2001lattice,shan1993lattice,zhang2005lattice,ambrus2019quadrature,chen2010multiple,li2012additional,wang2020simple,chen2018highly,wang2020simplified,saadat2020semi,
fei2019modeling,qiu2020study,qiu2021mesoscopic,sun2020discrete,sun2019anisotropic,zhan2021lattice,Huang2021transition,Wang2021Lattice} and the description method of non-equilibrium behavior in statistical physics. But it should be pointed out that, being different from the standard LBM in the usual sense, the DBM does not rely on the physical image of lattice gas model where the virtual particles propagate from current grid to an adjacent grid in one time step. In the absence of misunderstanding, DBM is also used as an abbreviation for discrete Boltzmann model or discrete Boltzmann method.

 As one of the specific applications of coarse-grained modeling theory in non-equilibrium statistical physics in the field of fluid mechanics, the DBM is a further development of phase space description method in the form of discrete Boltzmann equation\cite{Xu2021CJCP}. The methodology of DBM is to decompose complex problems into parts and select a perspective to study a set of kinetic properties of the system, so it requires that the kinetic moments describing this set of properties maintain their values in the model simplification. Based on the independent components of the kinetic moments of $(f - f^{eq})$, construct the phase space, and the phase space and its subspaces are used to describe the non-equilibrium behavior of the system. The research perspective and modeling accuracy will be adjusted as the research progresses. With the help of DBM, kinetic processes neglected by NS modeling, such as the non-equilibrium and mutual conversion of internal energy in different degrees of freedom during the reaction process, etc., can be investigated \cite{Xu2015APS,Xu-Chapter2,Xu2021AAS,Xu2021AAAS,Xu2021CJCP,Ji2021AIPA,Lin2020Entropy,Lin2018CAF,Lin2018CNF,ChenL2021FOP}, where $f$ and $f^{eq}$ are the distribution function and its corresponding equilibrium, respectively.

In order to simplify the problem and consistent with the existing literature, we focus mainly on a one-dimensional shock wave propagating in a fully ionized, quasi-neutral, homogeneous, unmagnetized plasma with no applied external electric and magnetic fields, and the ion-electron recombination effect are also neglected. The ion is in single species and the charge number $Z_i=1$. Moreover, the radiation effect is neglected, which means the only external force that charge particles subjected to is electric field force caused by charge separation.

This paper is arranged as follows. In section 2, we briefly introduce the DBM and the physical quantities used to describe plasma shock system. In section 3, we adopt two Riemann shock tube problems to verify the present DBM. In section 4, we introduce the calculation parameter settings. In section 5, we explore the ion peak structure and the corresponding TNE effects near the wavefront and give the relationship between ion peak distribution and the intensity of electric force. Then, we change the magnitude of $\rm{Ma}$ and investigate the variation trend of ion physical quantities and TNE effects near the peak. In section 6, we summarize and make a conclusion for the whole paper.

\section{Discrete Boltzmann modeling method}

The original Boltzmann equation reads
\begin{equation}\label{Eq:Boltzmann}
\frac{\partial f}{\partial t} + \boldsymbol{v} \cdot \frac{\partial f}{\partial \boldsymbol{r}} + \boldsymbol{a} \cdot \frac{\partial f}{\partial \boldsymbol{v}} = Q(f,f)
\end{equation}
where $f$ is the particle distribution function. The variables $t$ , $\boldsymbol{v}$ , $\boldsymbol{r}$ , $\boldsymbol{a}$, are the time, particle velocity, space coordinate, acceleration caused by external force, respectively. $Q(f,f)$ is the Boltzmann collision term, which represents the change of distribution function $f$ caused by particle collisions. From Boltzmann equation to discrete Boltzmann model, two steps of coarse-grained physical modeling are required. The principle of coarse-grained modeling is that the physical quantities we concern must keep the same values before and after simplification.

\subsection{Linearization of the collision term}

Compared with Liouville equation, Boltzmann equation is a much coarse-grained model. However, the collision term $Q(f,f)$ in the right-hand of Eq. \eqref{Eq:Boltzmann} is still too complicated to solve for most practical problems. A common practice is to substitute the collision term with a linearized collision operator, and write the Boltzmann equation as the following form,
\begin{equation}\label{Eq:Boltzmann-BGK}
\frac{\partial f}{\partial t} + \boldsymbol{v} \cdot \frac{\partial f}{\partial \boldsymbol{r}} + \boldsymbol{a} \cdot \frac{\partial f}{\partial \boldsymbol{v}} = -\frac{1}{\tau} (f - f^{eq})
\end{equation}
where $f^{eq}$ is the local particle equilibrium distribution function. The physical meaning of this practice is that $f$ evolves towards $f^{eq}$ through particle collisions, and the speed of this process is controlled by relaxation time $\tau$. According to the different forms of $f^{eq}$, the linearized collision model named as
Bhatnagar-Gross-Krook (BGK) model\cite{BGK1954-PR}, Ellipsoidal Statistical (ES) BGK model\cite{Holway1966New}, Shakov model (for monatomic gas)\cite{Shakhov1968}, Rykov model (for diatomic gas)\cite{Rykov2010}, etc. In this work, we adopt BGK model, where $f^{eq}$ is the Maxwellian distribution which reads
\begin{equation}\label{Eq:feq}
\begin{split}
& f^{eq} \left( \rho , \boldsymbol{u}  , T \right) =
\\ & \rho \left( \frac 1 {2 \pi RT} \right)^{D/2} \left[ \frac 1 {2 \pi nRT} \right]^{1/2} exp \left[ -\frac{(\boldsymbol{v}-\boldsymbol{u})^2}{2RT}-\frac{\eta^2}{2nRT} \right]
\end{split}
\end{equation}
where $\rho$, $\boldsymbol{u}$ and $T$ are the density, bulk velocity and temperature, respectively. $R$ is the gas constant. $D$ is the number of space dimension. $\eta$ represents the extra degree of freedom such as molecular rotation and vibration. $n$ is the number of extra degree of freedom, according to which the specific heat ratio $\gamma$ could be written as,
\begin{equation}\label{Eq:Specific-heat-ratio}
\gamma = (D+n+2)/(D+n)
\end{equation}
In the following discussion the mass of the particle (ion) described by the distribution function and the gas constant $R$ are assumed to be unity, and consequently the mass density equals to the number density and $T$ is used to replace $RT$.

\subsection{Discretization of the particle velocity space}

The discrete Boltzmann equation with discrete velocity $\boldsymbol{v}_i$  is as follows,
\begin{equation}\label{Eq:DBM-1}
\frac{\partial f_i}{\partial t} + {v_{i \alpha}} \frac{\partial f_i}{\partial {r_{\alpha}}} + \it{Force \,\, term} = -\frac1{\tau} (f_i - f_i^{eq})
\end{equation}
where the subscript $i$ ($= 1,..., N$) represents the $i_{th}$ discrete velocity, and the subscript $\alpha$ represents the $x$, $y$ and $z$ component of space in three-dimensional case.

$f(\boldsymbol{r},\boldsymbol{v},t)$ is defined in the $(6+1)$-dimensional phase space of particle position, particle velocity and time. Since particle can move towards any direction range from $(-\infty,+\infty)$, the conventional discrete method for discretizing time and space is not suitable for the discretization of particle velocity space. Therefore, the discrete distribution function $f_i$ do not have specific physical meaning, also do not correspond to the probability that particle velocity equal to $\boldsymbol{v}_i$. What we used to analyzing the system is not the specific value of $f_i$, but the kinetic moments. So the principle is that integral form of the kinetic moments is equal to the sum form, as follows,
\begin{equation}\label{Eq:DBM-principle1}
\int f \boldsymbol{\Psi(v)} d \boldsymbol{v} = \sum_{i} f_i \boldsymbol{\Psi(v_i)}
\end{equation}
where $\boldsymbol{\Psi(v)} = [1,\boldsymbol{v},\boldsymbol{vv},\cdots]$ correspond to different kinetic moments. It is easy to find that the dimension of $f_i$ equals to that of $f \boldsymbol{v}$. According to Chapman-Enskog multi-scale analysis, Eq. \eqref{Eq:DBM-principle1} can be finally expressed as
\begin{equation}\label{Eq:DBM-principle2}
\int f^{eq} \boldsymbol{\Psi'(v)} d \boldsymbol{v} = \sum_{i} f_i^{eq} \boldsymbol{\Psi'(v_i)}
\end{equation}
where $\boldsymbol{\Psi'(v)}$ corresponds to higher order kinetic moments.
A key step for the DBM simulation is the calculation of $f_i^{eq}$ on the right hand of Eq. \eqref{Eq:DBM-1}.

The modeling precision of DBM can be adjusted according to the practical needs.
As the first attempt for constructing DBM for plasma, in this paper we consider the case where the system deviates from its thermodynamic equilibrium slightly and consequently only the first order TNE effects are needed to be taken into account.

To construct such a DBM, the values of seven kinetic moments shown in the left hand side of Eqs. \eqref{Eq:Moment1}-\eqref{Eq:Moment7},
should keep unchanged when being calculated in the following summation form,
\begin{flalign}\label{Eq:Moment1}
& \int \int f^{eq} d\boldsymbol{v} d\boldsymbol{\eta} = \rho = \sum{f_i^{eq}}, &
\end{flalign}
\begin{flalign}\label{Eq:Moment2}
& \int \int f^{eq} \boldsymbol{v} d\boldsymbol{v} d\boldsymbol{\eta} = \rho \boldsymbol{u} = \sum{f_i^{eq}\boldsymbol{v}}, &
\end{flalign}
\begin{flalign}\label{Eq:Moment3}
&\int \int  f^{eq} \frac{1}{2} \left(v^2 + \eta^2 \right) d\boldsymbol{v} d\boldsymbol{\eta} = E_T &  \\ \nonumber
&= \rho \left[ \frac{\left( D + n \right) T}{2} + \frac {u^2}{2} \right] = \sum{f_i^{eq} \frac{1}{2} \left( v_i^2 + \eta_i^2 \right)},& \nonumber
\end{flalign}
\begin{flalign}\label{Eq:Moment4}
& \int \int f^{eq} \boldsymbol{v} \boldsymbol{v} d\boldsymbol{v} d\boldsymbol{\eta} = p \boldsymbol{I} + \rho \boldsymbol{u} \boldsymbol{u} = \sum{f_i^{eq}\boldsymbol{v}_i\boldsymbol{v}_i}, &
\end{flalign}
\begin{flalign}\label{Eq:Moment5}
& \int \int f^{eq} \frac{1}{2} \left(v^2 + \eta^2 \right) \boldsymbol{v} d\boldsymbol{v} d\boldsymbol{\eta} = \left( E_T + p \right) \boldsymbol{u} = \rho \boldsymbol{u}& \\ \nonumber
& \left[ \frac{\left( D + n + 2 \right) T}{2} + \frac {u^2}{2} \right] = \sum{f_i^{eq} \frac{1}{2} \left( v_i^2 + \eta_i^2 \right) \boldsymbol{v}_i},& \nonumber
\end{flalign}
\begin{flalign}\label{Eq:Moment6}
& \int \int f^{eq} \boldsymbol{v} \boldsymbol{v} \boldsymbol{v} d\boldsymbol{v} d\boldsymbol{\eta} = p ( \boldsymbol{u_{\alpha}} \boldsymbol{e_{\beta}} \boldsymbol{e_{\gamma}} \delta_{\beta \gamma} +
\boldsymbol{e_{\alpha}} \boldsymbol{u_{\beta}} \boldsymbol{e_{\gamma}}& \\ \nonumber
& \delta_{\alpha \gamma} + \boldsymbol{e_{\alpha}} \boldsymbol{e_{\beta}} \boldsymbol{u_{\gamma}} \delta_{\alpha \beta} )+ \rho \boldsymbol{u} \boldsymbol{u} \boldsymbol{u} = \sum{f_i^{eq} \boldsymbol{v}_i \boldsymbol{v}_i \boldsymbol{v}_i},& \nonumber
\end{flalign}
\begin{flalign}\label{Eq:Moment7}
& \int \int f^{eq} \frac{1}{2} \left(v^2 + \eta^2 \right) \boldsymbol{v} \boldsymbol{v} d\boldsymbol{v} d\boldsymbol{\eta} = T \left( E_T + p \right) \boldsymbol{I}& \\ \nonumber
& + \boldsymbol{u} \boldsymbol{u} \left( E_T + 2p \right) = p \left[ \frac {\left( D + n + 2 \right) T}{2} + \frac{u^2}{2} \right]&  \\ \nonumber
& \boldsymbol{I} + \rho \boldsymbol{u} \boldsymbol{u} \left[ \frac {\left( D + n + 4 \right) T}{2} + \frac{u^2}{2} \right] = &\\ \nonumber
& \sum{f_i^{eq} \frac{1}{2} \left( v_i^2 + \eta_i^2 \right) \boldsymbol{v}_i \boldsymbol{v}_i},& \nonumber
\end{flalign}
where
$\eta_i$ is the discrete correspondence of $\eta$. The DBM for higher order TNE flows can be constructed in a similar way via considering higher order TNE effects (i.e. more kinetic moments have to be considered).

In this work, we consider only one type of ion with charge $Z_i = 1$. The electron is assumed to be inertialess and always in thermodynamic equilibrium. The behavior of ion is described by the distribution function $f$ which follows Eq. \eqref{Eq:DBM-1}.
The only interaction taken into account between ion and electron is the electric field force caused by charge separation.
The electric field force $\boldsymbol{a}=e \boldsymbol{E}$ is added into the forcing term, where $e$ is the charge of proton.
When the non-equilibrium effects caused by $f$ deviating from $f^{eq}$ in force term is not significant, we have
\begin{equation}\label{Eq:Force-term}
\boldsymbol{a} \cdot \frac{\partial f}{\partial \boldsymbol{v}} \approx \boldsymbol{a} \cdot \frac{\partial f^{eq}}{\partial \boldsymbol{v}} = - e\boldsymbol{E} \cdot \frac{(\boldsymbol{v} - \boldsymbol{u})}{T} f^{eq}.
\end{equation}
Consequently, the DBM evolution equation \eqref{Eq:DBM-1} becomes
\begin{equation}\label{Eq:DBM-f0}
\frac{\partial f_i}{\partial t} + {v_{i \alpha}} \frac{\partial f_i}{\partial {r_{\alpha}}}
- e\boldsymbol{E} \cdot \frac{(\boldsymbol{v} - \boldsymbol{u})}{T} f_i^{eq}
= -\frac1{\tau} (f_i - f_i^{eq}).
\end{equation}
It is easy to find that the following hydrodynamic model,
\begin{equation}\label{Eq:NS}
\left\{
\begin{aligned}
& \frac{\partial \rho}{\partial t} + \nabla \cdot (\rho \boldsymbol{u}) = 0 ,\\
& \frac{\partial \rho \boldsymbol{u}}{\partial t} + \nabla \cdot (\rho \boldsymbol{uu}) + \nabla p = - \nabla \cdot \boldsymbol{P'} +\rho e\boldsymbol{E} ,\\
& \frac{\partial E_{T}}{\partial t} + \nabla \cdot \left[ (E_{T}+p)\boldsymbol{u} \right] = \nabla \cdot \left[ \kappa \nabla T + \boldsymbol{P'} \cdot \boldsymbol{u} \right] + \rho e\boldsymbol{E} \cdot \boldsymbol{u},\\
\end{aligned}
\right.
\end{equation}
is one part of the current DBM,
where $p=\rho T$ is the pressure, $E_{T} = \rho e_{int} + (\rho u^{2})/2$ is the system energy per unit volume, and $e_{int}=(n+2)T/2$ is the internal energy. $\mu = \tau p $ and $\kappa = c_p \tau p $ are the coefficients of viscosity and thermal conductivity, respectively.

In addition to the HNE behavior, DBM could simultaneously give the the most relevant TNE effects accompanying with the macroscopic flows. In other words, DBM can be regarded as a hydrodynamic model supplemented by a coarse-grained model of the most relevant TNE effects. The way to obtain TNE quantities is by comparing the distribution function $f$ to the local equilibrium distribution function $f^{eq}$ with kinetic moments at a certain time, which defined as,
\begin{equation}\label{Eq:TNE1}
\boldsymbol{\Delta}_{m} = \boldsymbol{M}_{m}(f)-\boldsymbol{M}_{m}(f^{eq}),
\end{equation}
\begin{equation}\label{Eq:TNE2}
\boldsymbol{\Delta}^*_m = \boldsymbol{M^*}_{m}(f)-\boldsymbol{M}^{*}_{m}(f^{eq}),
\end{equation}
where $\boldsymbol{M}_{m}$ represent the different orders of non-central kinetic moments involving bulk velocity, and $\boldsymbol{M}^{*}_{m}$ represents the different orders of central kinetic moments describing the thermal fluctuation information. Of all the kinetic moments, the first three will always be equal when displace $f^{eq}$ with $f$, which is determined by the conservation law of mass, momentum and energy. However, the result will be different for higher order moments, and each of these non-equilibrium quantities reflects the extent system deviated from local thermodynamic equilibrium from one aspect. The non-equilibrium quantities used in this paper are defined as follows,
\begin{equation}\label{Eq:delta}
\left\{
\begin{aligned}
& \boldsymbol{\Delta}_2 = \sum{f_i \boldsymbol{v}_i \boldsymbol{v}_i} - \sum{f_i^{eq} \boldsymbol{v}_i \boldsymbol{v}_i} ,\\
& \boldsymbol{\Delta}_{3,1} = \sum{f_i \left( v_i^2 + \eta_i^2 \right) \boldsymbol{v}_i} - \sum{f_i^{eq} \left( v_i^2 + \eta_i^2 \right) \boldsymbol{v}_i} ,\\
& \boldsymbol{\Delta}_3 = \sum{f_i \boldsymbol{v}_i \boldsymbol{v}_i \boldsymbol{v}_i} - \sum{f_i^{eq} \boldsymbol{v}_i \boldsymbol{v}_i \boldsymbol{v}_i}   ,\\
& \boldsymbol{\Delta}_{4,2} = \sum{f_i \left( v_i^2 + \eta_i^2 \right) \boldsymbol{v}_i \boldsymbol{v}_i} - \sum{f_i^{eq} \left( v_i^2 + \eta_i^2 \right) \boldsymbol{v}_i \boldsymbol{v}_i} ,\\
\end{aligned}
\right.
\end{equation}
By substituting $\boldsymbol{v}_i$ in Eq. \eqref{Eq:delta} with $(\boldsymbol{v}_i - \boldsymbol{u})$, a TNE quantity $\boldsymbol{\Delta}^*_m$ based on the central kinetic moments can also be obtained.

In this work we consider a one-dimensional shock, so the electric field force is only in $x$ direction. Thus, we get the following evolution equation,
\begin{equation}\label{Eq:DBM-f}
\frac{\partial f_i}{\partial t} + {v_{i \alpha}} \frac{\partial f_i}{\partial {r_{\alpha}}} - \frac{e E_x (v_{i x} - u_{x})}{T} f_i^{eq} = -\frac1{\tau} (f_i - f_i^{eq}).
\end{equation}
The Poisson equation gives
\begin{equation}\label{Eq:Poisson}
\frac{{{d^2}\varphi }}{{d{x^2}}} =  - \frac{{e\left( {{n_i} - {n_e}} \right)}}{{{\varepsilon _0}}}
\end{equation}
where $\varphi$ is the space potential, $n_i$ is the ion number density which is equals to $\rho_i$ and $n_e$ is the electron number density. $\varepsilon _0$ is the vacuum permittivity. The electron is assumed to be always in thermodynamic equilibrium, so the electron density obeys Boltzmann distribution as
\begin{equation}\label{Eq:Boltzmann distribution}
{n_e} = {n_{e0}}\exp \left( {\frac{{e\varphi }}{{k{T_e}}}} \right)
\end{equation}
Where $n_{e0}$ is the initial electron density when $\varphi$ is equal to zero. $k=R/N_A$ is the Boltzmann constant, and is equal to $1$ for particle mass $N_A$ and gas constant $R$ are unity.  By substituting Eq. \eqref{Eq:Poisson} with Eq. \eqref{Eq:Boltzmann distribution} and assuming the vacuum permittivity $\varepsilon _0$ is unity, the Poisson equation reads
\begin{equation}\label{Eq:Poisson2}
\frac{{{d^2}\varphi }}{{d{x^2}}}  = e{n_{e0}}\exp \left( {\frac{{e\varphi }}{{{T_e}}}} \right) - e{n_i}
\end{equation}
The proton charge $e$, electron temperature $T_e$, and initial electron density $n_{e0}$ are assumed to be unity.
Then the Poisson reads
\begin{equation}\label{Eq:Poisson3}
\frac{{{d^2}\varphi}}{{d{x^{2}}}} = \exp \left( {\varphi} \right) - {n_i}
\end{equation}

\begin{figure}
\centering
\includegraphics[height=6cm]{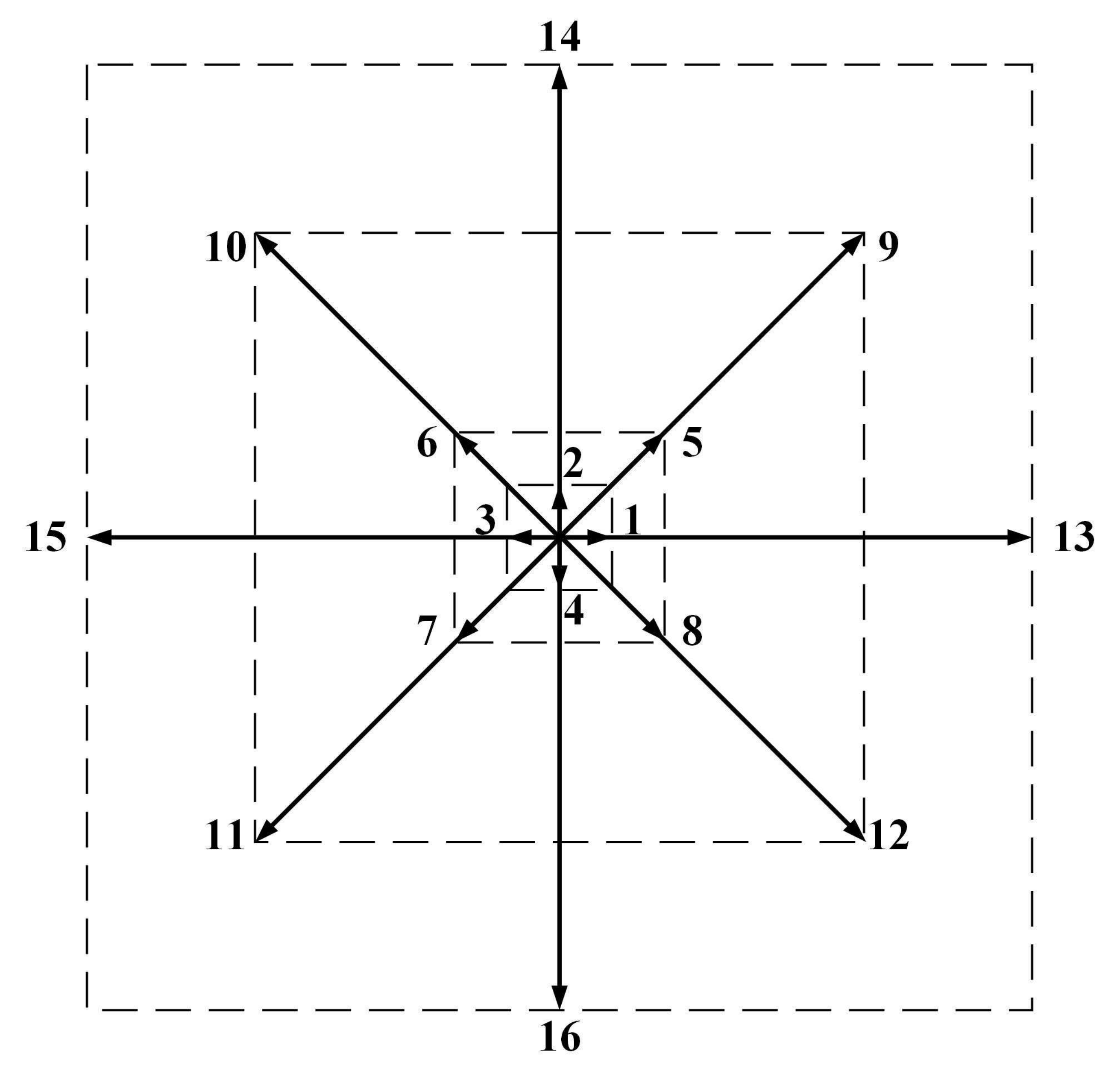}
\caption{Schematic of the discrete velocity model.}
\label{Fig:DVM}
\end{figure}

In this work, we consider only the two dimensional case.  The seven kinetic moment relations, \eqref{Eq:Moment1}-\eqref{Eq:Moment7}, have sixteen components in two-dimensional case, so at least sixteen discrete velocities are needed. The sketch map of discrete velocity model we adopted is shown in Figure \ref{Fig:DVM}.
The values of discrete velocities are shown in Eq. \eqref{Eq:value_DVM},
\begin{equation}\label{Eq:value_DVM}
\boldsymbol{v}_i = \left\{
\begin{aligned}
& v_1         \left[  \cos  \frac  {(i-1)\pi}{2},   \sin  \frac  {(i-1)\pi}{2}   \right], & i = 1-4    \\
& \sqrt{2}v_2 \left[  \cos  \frac  {(2i-1)\pi}{4},  \sin  \frac  {(2i-1)\pi}{4}  \right], & i = 5-8    \\
& \sqrt{2}v_3 \left[  \cos  \frac  {(2i-9)\pi}{4},  \sin  \frac  {(2i-9)\pi}{4}  \right], & i = 9-12   \\
& v_4         \left[  \cos  \frac  {(i-13)\pi}{2},  \sin  \frac  {(i-13)\pi}{2}  \right], & i = 13-16  \\
\end{aligned}
\right.
\end{equation}
and $\eta_i =\eta_0$ for $i = 5,6,7,8$, else $\eta_i=0$.

\section{Validation and verification}

In this section, we choose two typical one-dimensional Riemann problems to confirm the validity of the present model for capturing main structures in flow. And the spatial and temporal derivatives are discretized by using forward Euler finite difference scheme and the nonoscillatory nonfree dissipative (NND) scheme\cite{zhang1991nnd}, respectively. By using these two discretization schemes, the simulation has achieved 1st-order accuracy in time and 2nd-order accuracy in space, respectively.

\subsection{Sod shock tube problem}

\begin{figure}
\centering
\includegraphics[height=7cm]{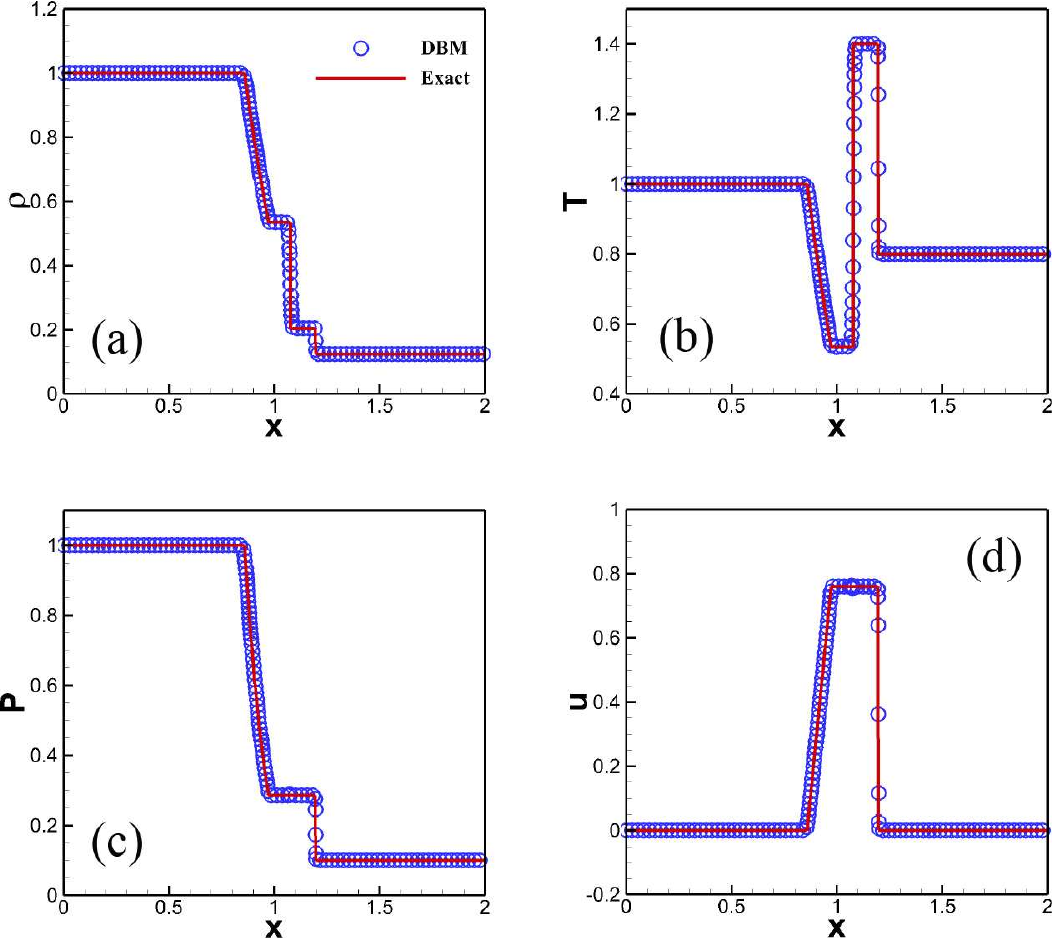}
\caption{Comparison between Riemann solutions and DBM results of Sod shock tube problem. (a) density, (b) pressure, (c) temperature, (d) horizontal velocity.}
\label{Fig:Sod}
\end{figure}

The initial conditions are as follows,
\begin{equation}
\left\{
\begin{aligned}
(&\rho,T,u,v)_{L} = (1.0,1.0,0.0,0.0) ,& x\leq1 \\
(&\rho,T,u,v)_{R} = (0.125,0.8,0.0,0.0), & x>1
\end{aligned}
\right.
\end{equation}
where the subscript ``L" and ``R" represent the left and right sides of the discontinuity interface. The grid number of calculated region is $[N_x \times N_y] = [2000\times 1]$, and the initial interface is located at $x=1$. The meshing size is $\Delta x = \Delta y = 1\times 10^{-3}$, and the time step is $\Delta t = 1\times 10^{-5}$. Parameters for discrete velocities are $v_1 = 0.5$, $v_2 = 1.0$, $v_3 = 2.9$, and $v_4 = 4.5$. The parameter for extra degree is $\eta_0 = 5$. The other parameters are $\tau = 2\times 10^{-5}$, $n = 0 \left(i.e., \gamma = 2 \right)$. In the $y$ direction, the periodic boundary is adopted. In the $x$ direction, the left and right boundary are assumed always in the initial equilibrium state, which are
\begin{equation}
\left\{
\begin{aligned}
& f_{i,-1,t} = f_{i,0,t} = f^{eq}_{i,1,t = 0} \\
& f_{i,N_x +2,t} = f_{i,N_x +1,t} = f^{eq}_{i,N_x,t = 0}
\end{aligned}
\right.
\end{equation}
where the subscripts ``$-1$", ``$0$", ``$N_x +1$" and ``$N_x +2$" represent the ghost nodes on the left and right sides. Figure \ref{Fig:Sod} shows the comparison between the analytical result of sod shock tube problem and the result of DBM at $t = 0.1$. Obviously, the result of DBM is in good agreement with the exact solution. Here the exact solution is in Euler level, which means the dissipation effects such as viscosity and thermal conduction have been ignored.

\subsection{Lax shock tube problem}

\begin{figure}
\centering
\includegraphics[height=7cm]{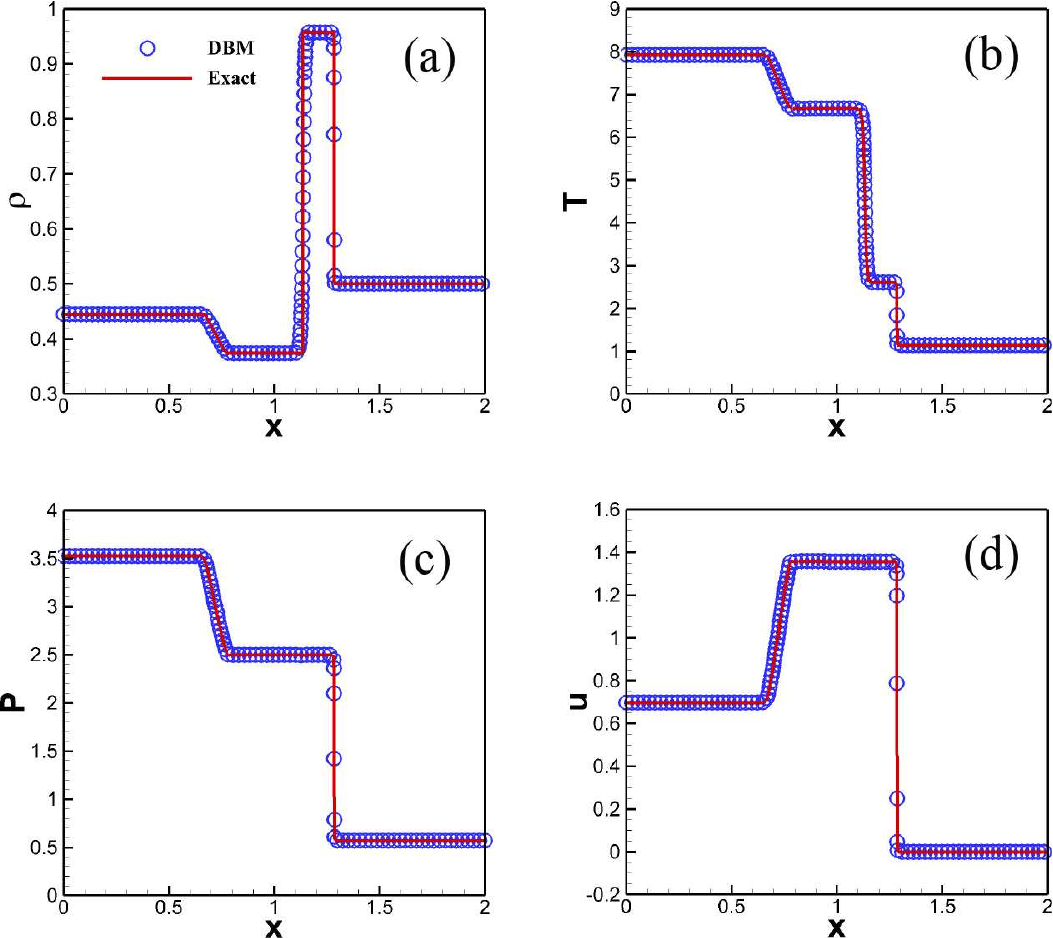}
\caption{Comparison between Riemann solutions and DBM results of Lax shock tube problem. (a) density, (b) pressure, (c) temperature, (d) horizontal velocity.}
\label{Fig:Lax}
\end{figure}

The initial conditions are as follows,
\begin{equation}
\left\{
\begin{aligned}
(&\rho,T,u,v)_{L} = (0.445,7.928,0.698,0.0) ,& x\leq1 \\
(&\rho,T,u,v)_{R} = (0.5,1.142,0.0,0.0) ,& x>1
\end{aligned}
\right.
\end{equation}
Figure \ref{Fig:Lax} shows the computation result of density, pressure, temperature and velocity in $x$ direction at $t = 0.1$, where the blue circles indicate the DBM results and the red solid lines indicate the exact Riemann solution in Euler level. The grid number of calculated region is $[N_x \times N_y] = [2000\times 1]$, and the initial interface is also located at $x=1$. The parameter are set to be $\Delta x = \Delta y = 1\times 10^{-3}$, $\Delta t = 1\times 10^{-5}$, $\tau = 2\times 10^{-5}$, $\eta_0 = 5$, $n = 0 \left(i.e., \gamma = 2 \right)$, and $v_1 = 0.5$, $v_2 = 1.0$, $v_3 = 2.9$, $v_4 = 4.5$. The boundary conditions in $x$ and $y$ direction are consistent with the setting in sod shock tube problem. From Figure \ref{Fig:Lax}, we can observe that the two results are in good agreement with each other.

From the results of two one-dimensional Riemann problems, we find that the present DBM with appropriate discretization schemes can capture the main structure of flow with shock wave, expanding wave and contact discontinuity effectively, which is a basic capability for simulating shock wave propagating in plasma.

\section{Calculation parameter settings}

In the simulation of plasma shock wave, the initial macroscopic quantities are arranged as follows,
\begin{equation}
\left\{
\begin{aligned}
(&\rho,u_1,u_2,T)_{L} = (\rho_0,u_0,0,T_0), & x \leq N_x/8 \\
(&\rho,u_1,u_2,T)_{R} = (1.0,0.0,0.0,1.0), & x > N_x/8
\end{aligned}
\right.
\end{equation}
where the subscripts ``L" and ``R" indicates the downstream and upstream of shock wave. $\rho_0$, $u_0$, $T_0$ are the initial downstream density, bulk velocity, temperature, respectively. The initial upstream and downstream macroscopic quantities are connected by the Rankine-Hugoniot relations, which can be deduced from Eq. \eqref{Eq:NS} after several steps. First, by substituting Eq. \eqref{Eq:NS} with Eq. \eqref{Eq:Poisson3} we get the following equations
\begin{equation}\label{Eq:NS2}
\left\{
\begin{aligned}
& \frac{\partial \rho}{\partial t} + \nabla \cdot (\rho \boldsymbol{u}) = 0 ,\\
& \frac{\partial \rho \boldsymbol{u}}{\partial t} + \nabla \cdot (\rho \boldsymbol{uu}) + \nabla p = \left[ {{\nabla ^2}\varphi  - \exp(\varphi) } \right]\nabla \varphi  ,\\
& \frac{\partial E_{T}}{\partial t} + \nabla \cdot \left[ (E_{T}+p)\boldsymbol{u} \right] = - \rho u\nabla \varphi,\\
\end{aligned}
\right.
\end{equation}
After simplification, Eq. \eqref{Eq:NS2} becomes
\begin{equation}\label{Eq:NS3}
\left\{
\begin{aligned}
& \frac{\partial \rho}{\partial t} + \nabla \cdot (\rho \boldsymbol{u}) = 0 ,\\
& \frac{\partial \rho \boldsymbol{u}}{\partial t} + \nabla \cdot [\rho \boldsymbol{uu} + p - {\frac{{{{\left[ {\nabla \varphi } \right]}^2}}}{2} + \exp(\varphi) }] = 0  ,\\
& \frac{\partial E_{T}}{\partial t} + \varphi \frac{\partial \rho}{\partial t} + \nabla \cdot \left[ (E_{T} + p +\rho \varphi)\boldsymbol{u} \right] = 0,\\
\end{aligned}
\right.
\end{equation}
For the steady state, Eq. \eqref{Eq:NS3} becomes
\begin{equation}\label{Eq:NS4}
\left\{
\begin{aligned}
& \nabla \cdot (\rho \boldsymbol{u}) = 0 ,\\
& \nabla \cdot [\rho \boldsymbol{uu} + p - {\frac{{{{\left( {\nabla \varphi } \right)}^2}}}{2} + \exp(\varphi) }] = 0  ,\\
& \nabla \cdot \left[ (E_{T} + p +\rho \varphi)\boldsymbol{u} \right] = 0,\\
\end{aligned}
\right.
\end{equation}
For the steady state upstream and downstream flow, there exist no electric current and charge separation, which means $\nabla \varphi$ equals to zero. Then the Rankine-Hugoniot relations
\begin{equation}\label{Eq:relation}
\left\{
\begin{aligned}
& {\rho _L}{u_L} = {\rho _R}{u_R} \\
& {\rho _L}{u_L}{u_L} + {\rho _L}{T_L} + \exp ({\varphi _L}) = {\rho _R}{u_R}{u_R} + {\rho _R}{T_R} + \exp ({\varphi _R}) \\
& \frac{\gamma }{{\gamma  - 1}}{T_L} + \frac{{u_L^R}}{R} + {\varphi _L} = \frac{\gamma }{{\gamma  - 1}}{T_R} + \frac{{u_R^2}}{2} + {\varphi _R}
\end{aligned}
\right.
\end{equation}
are deduced. Also, the Poisson equation becomes
\begin{equation}\label{Eq:Poisson4}
\rho_i  = n_i = \exp(\varphi)
\end{equation}
After setting the initial conditions, the the Poisson equation is calculated for the whole computational domain with time evolution. The accurate solution of Poisson is assumed as follows
\begin{equation}\label{Eq:Accurate Poisson}
\varphi  = {\varphi _0} + \delta \varphi
\end{equation}
Where ${\varphi _0}$ is the initial value of potential, and $\delta \varphi$ is the deviation of ${\varphi _0}$ from the exact value $\varphi$. By inserting Eq. \eqref{Eq:Accurate Poisson} into Eq. \eqref{Eq:Poisson3}, the Poisson equation reads
\begin{equation}\label{Eq:Poisson5}
{\partial ^2}_x({\varphi _0}) + {\partial ^2}_x(\delta \varphi ) - \exp ({\varphi _0} + \delta \varphi ) + {n_i} = 0
\end{equation}
It is further assumed that
\begin{equation}\label{Eq:Poisson6}
{\partial ^2}_x({\varphi _0}) - \exp ({\varphi _0}) + {n_i} = {f_0}
\end{equation}
Thus the Poisson equation reads,
\begin{equation}\label{Eq:Poisson7}
{\partial ^2}_x(\delta \varphi ) - \exp ({\varphi _0})\delta \varphi  =  - {f_0}
\end{equation}
Eq. \eqref{Eq:Poisson7} is a tridiagonal matrix, so the chase method is used for calculation. The boundary condition for solving Eq. \eqref{Eq:Poisson7} is $\delta \varphi^1 = 0$ and $\delta \varphi^N = 0$, where ``1" and ``N" represent the left and right boundary.
The grid number of calculated region is $[N_x \times N_y] = [10000\times 1]$, and the initial interface is located at $N_x /8$. The simulation conditions are $\Delta x = \Delta y = 5\times 10^{-3}$, $\Delta t = 1\times 10^{-4}$, $\tau = 2\times 10^{-4}$, $n = 0 \left(i.e., \gamma = 2 \right)$. In order to maintain the stability of model, we choose $v_1 = 0.5$, $v_2 = 1.0$, $v_3 = 2.9$, $v_4 = 4.5$, and $\eta_0 = 5$.

\section{Result and discussion}

\subsection{Macroscopic quantities around plasma shock wave}
\begin{figure}
\centering
\includegraphics[height=3.5cm]{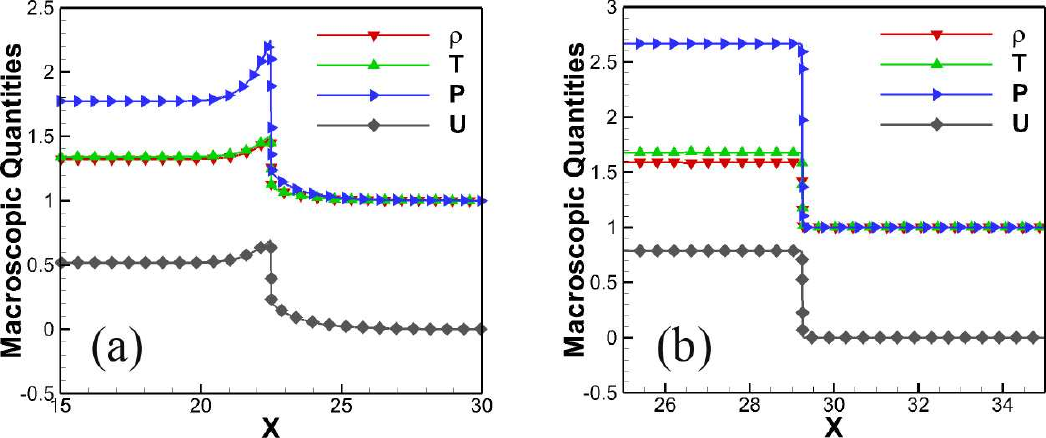}
\caption{Density, temperature, pressure and horizonal velocity distribution when $\rm{Ma}=1.5$. (a) plasma shock wave at $t=8$, (b) shock wave in neutral fluids at $t=2$.}
\label{Fig:Ma1.5}
\end{figure}
\begin{figure}
\centering
\includegraphics[height=3.5cm]{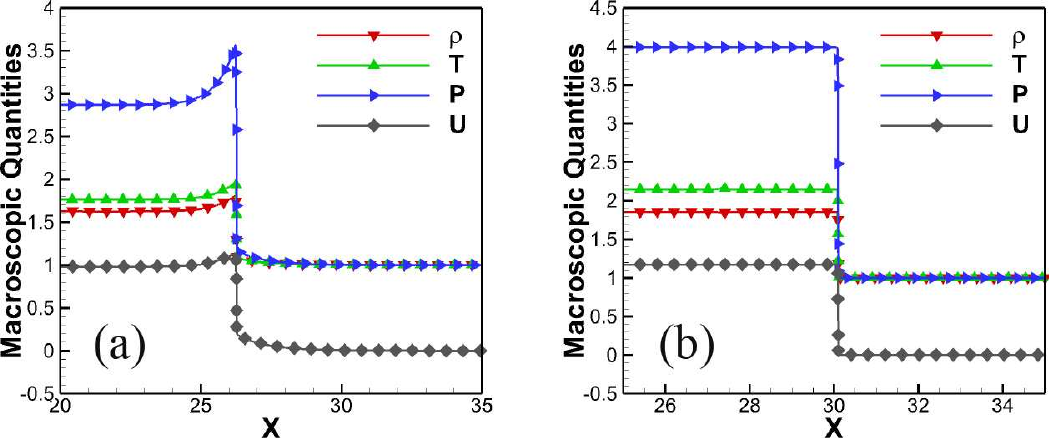}
\caption{Density, temperature, pressure and horizonal velocity distribution when $\rm{Ma}=1.8$. (a) plasma shock wave at $t=8$, (b) shock wave in neutral fluids at $t=2$.}
\label{Fig:Ma2.0}
\end{figure}
\begin{figure}
\centering
\includegraphics[height=3.5cm]{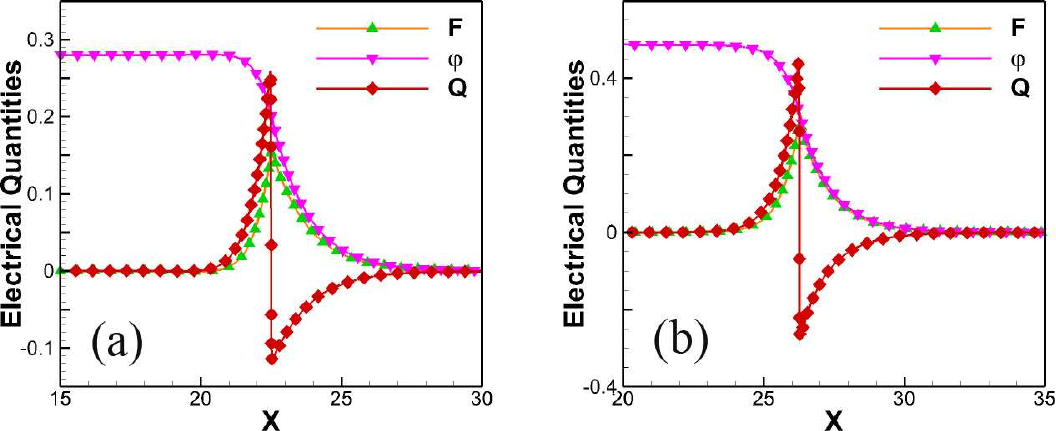}
\caption{Electric field force, potential and net charge distribution of plasma shock wave at $t=8$. (a) $\rm{Ma}=1.5$, (b) $\rm{Ma}=1.8$.}
\label{Fig:Ele1518}
\end{figure}

We first investigate the steady state structure of plasma shock wave. Figures \ref{Fig:Ma1.5} and \ref{Fig:Ma2.0} give the macroscopic quantities, including density, pressure, temperature and velocity in $x$ direction, around plasma and neutral fluid shock when $\rm{Ma}=1.5$ and $\rm{Ma}=1.8$, respectively. It is found that the plasma shock wave is very different from the shock wave in neutral fluids, and somewhat similar to the detonation wave. The macroscopic quantities both exhibit spike structures and reach the maximum value in the same position, but the maximum value of these macroscopic quantities are all less than the corresponding downstream value of shock wave in neutral fluids.
For neutral fluids, dissipation mechanisms such as viscosity and heat conduction near the wave front tend to make the upstream and downstream macroscopic quantities of the shock wave continuously distributed near the wave front. For plasma shock waves, since we only consider the effect of electric field force on ions, and the electric field force pushes ions from downstream to upstream. Consequently, the temperature, velocity, and pressure of the ions increase. At the same time, the dissipation mechanism in the ions works. Although it is not enough to smooth out the influence of the electric field force on the ions, the dissipation eventually makes the macroscopic quantities near the wavefront tend to be continuous, forming a spike structure which is different from the neutral shock wave.
Besides, the result is also different comparing with previous studies for not only temperature but also density, velocity and pressure appear maximum value that exceeds the downstream equilibrium value. The reason is that the exchange of momentum and energy between electrons and ions are ignored in our hypothesis.

Figure \ref{Fig:Ele1518} describes the electrical quantities, including electric field force, potential and net charge distribution in $x$ direction, around plasma when $\rm{Ma}=1.5$ and $\rm{Ma}=1.8$, respectively. It is observed that the electric field force also behave as a spike, but the net charge presenting as two opposite spikes. Through further analysis of data, it is found that the peak position of electric field force does not coincide with macroscopic quantities, but locate at the position where the net charge $Q=0$. The peak position of positive net charge is coincide with macroscopic quantities, but the peak position of negative net charge is locate at upstream. Intuitively, the net charge represents the extent of charge separation. Because the proton charge $e$ is assume to unity, the net charge is also equal to net density, so there occurs charge separation or density difference during the motion of plasma shock wave and forms the ion and electron concentration region in the downstream and upstream, respectively. In terms of the net charge spikes amplitude and width, it is observed that the absolute value of positive net charge peak is greater than that of negative net charge peak, but the negative net charge region is wider than positive net charge region, which is determined by the fact that the total charge of plasma is zero. The distribution of net charge also demonstrate that the electron tends to move towards upstream and the ion tends to move towards downstream, and the diffusion area of electron is longer than ion because of the small electron/ion mass ratio.

\subsection{Macroscopic and electrical quantities around plasma shock wave}
\begin{figure}
\centering
\includegraphics[height=3.5cm]{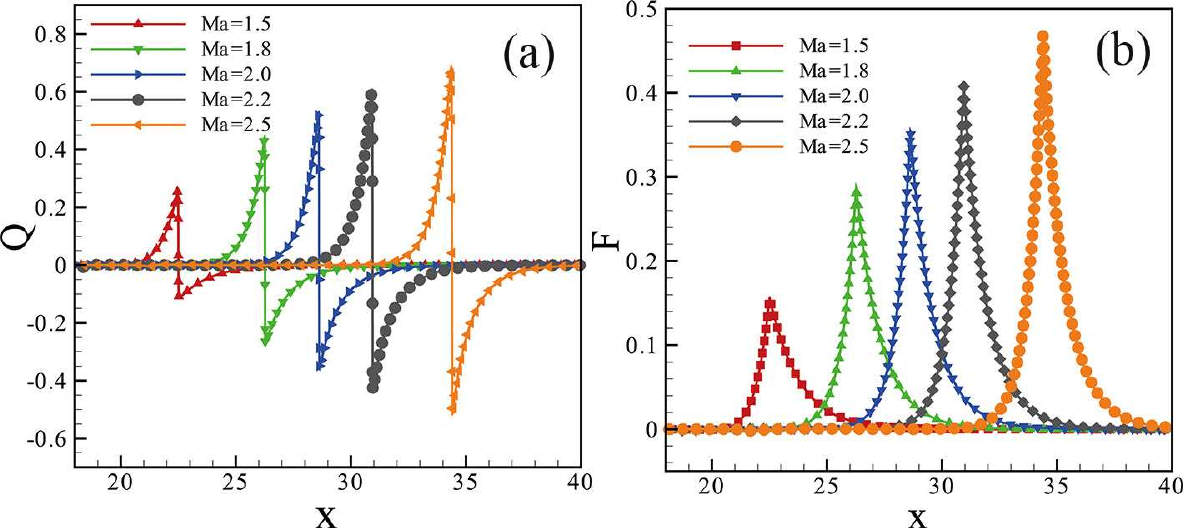}
\caption{Quantity of charge and electric field force varies with $\rm{Ma}$ at $t=8$. (a) charge, (b) electric field force.}
\label{Fig:Charge}
\end{figure}
We then investigate the variations of macroscopic and electrical quantities with Ma number. Figure \ref{Fig:Charge} shows the net charge and electric field force distribution around the shock front for the cases of $\rm{Ma} = 1.5,1.8,2.0,2.2,2.5$. It is found that an electric double layer appeared around the shock front. The large inertia of the ions causes them to lag behind, so the wave front charge is negative and the wave rear charge is positive. The peak value on the left side is larger than that on the right side, and the width on the left is smaller than the right, which means the left side is much steeper than the right side. As the Mach number increase, the two peaks of both two sides increase, indicating that electrons tend to move upstream of the shock wave with the increasing of $\rm{Ma}$. This movement tendency makes the degree of charge separation increase, and the electric field force also becomes larger.

\begin{figure}
\centering
\includegraphics[height=7.5cm]{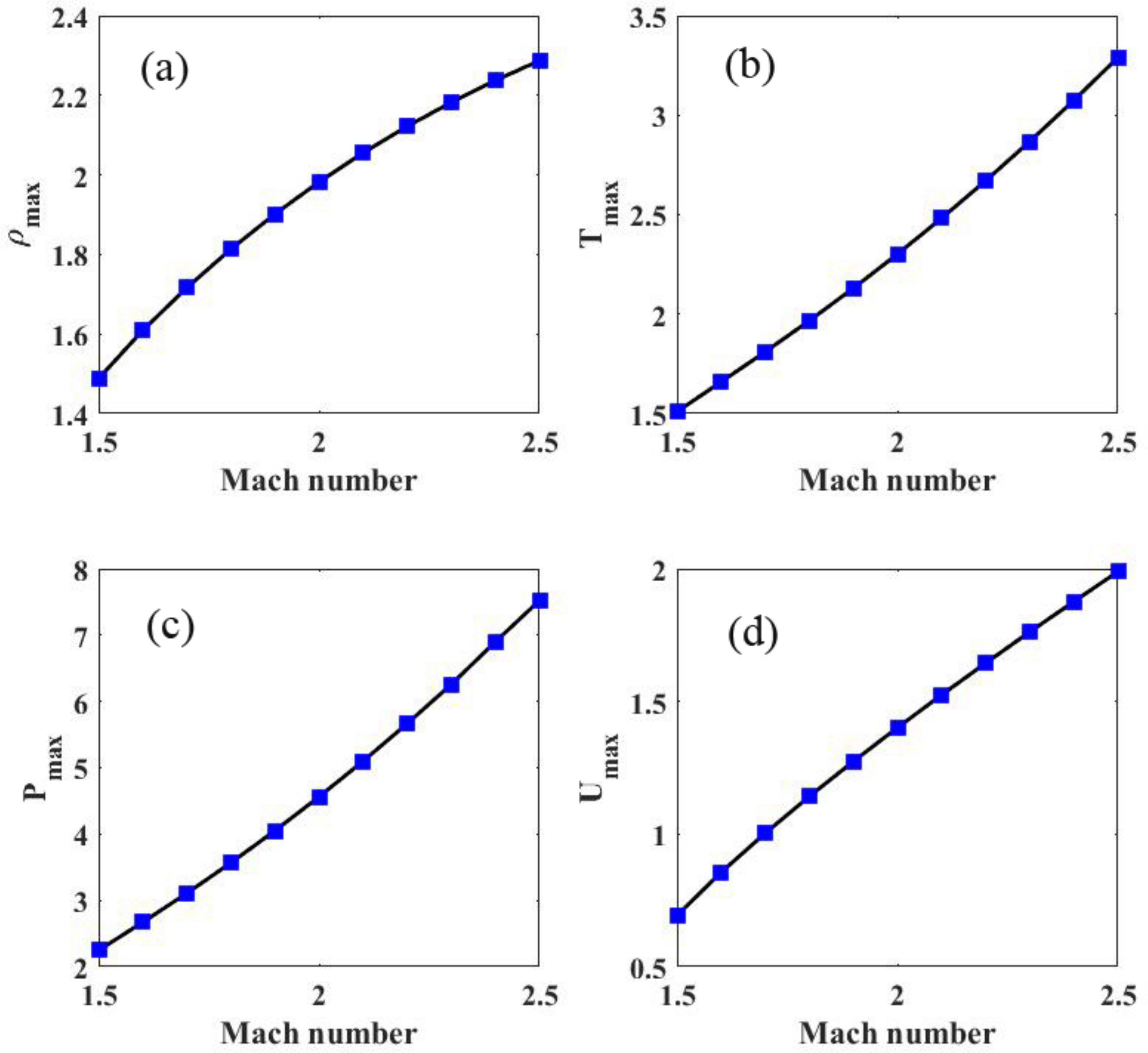}
\caption{Macroscopic quantities peak value varies with $\rm{Ma}$. (a) density, (b) temperature, (c) pressure, (d) horizontal velocity.}
\label{Fig:Macro}
\end{figure}
\begin{figure}
\centering
\includegraphics[height=7.5cm]{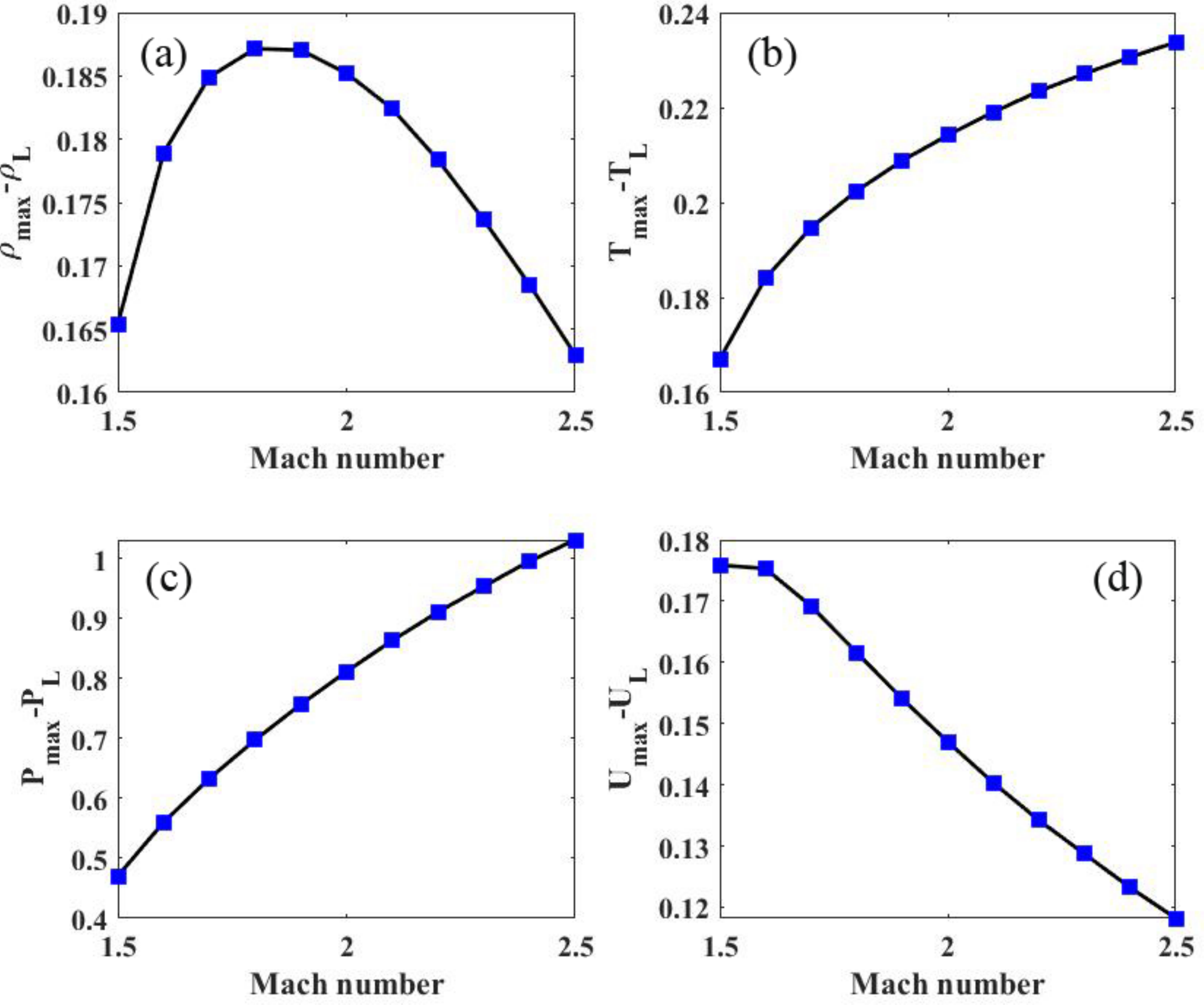}
\caption{Difference value between macroscopic quantities peak value and downstream value varies with $\rm{Ma}$. (a) density, (b) temperature, (c) pressure, (d) horizontal velocity.}
\label{Fig:Dif}
\end{figure}
\begin{figure}
\centering
\includegraphics[height=7.5cm]{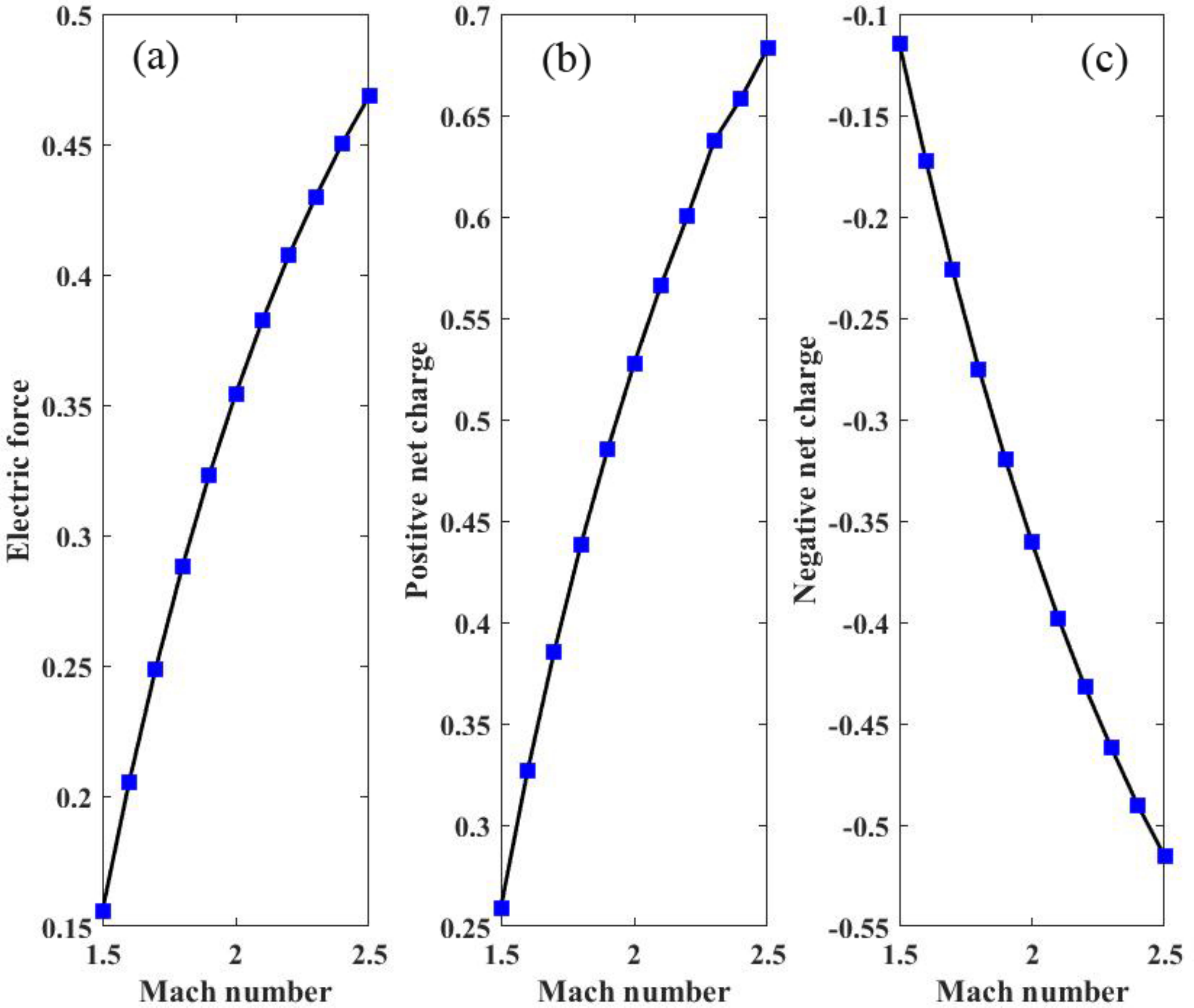}
\caption{Electrical quantities peak value varies with $\rm{Ma}$. (a) electric force, (b) positive net charge, (c) negative net charge.}
\label{Fig:Ele3}
\end{figure}
Figure \ref{Fig:Macro} shows the peak values of macroscopic quantities varies with $\rm{Ma}$. Obviously, the four peak values increases with $\rm{Ma}$ approximately in linear form. However, the growth rate of temperature and pressure increase slowly with $\rm{Ma}$, and the growth rate of density and horizontal velocity decrease slowly with $\rm{Ma}$.

Figure \ref{Fig:Dif} gives the evolution of differences between macroscopic quantity peak values and the corresponding downstream values with increasing $\rm{Ma}$.
It is observed that the difference of density first increases, then decreases with $\rm{Ma}$. The differences of temperature and pressure increase with $\rm{Ma}$. The difference of velocity decreases linearly with $\rm{Ma}$.
From Figures \ref{Fig:Dif} (b) and (c), it is also observed that the growth rates of both decrease gradually with the increase of $\rm{Ma}$, and the growth rate of temperature decreases faster.
When ions move from upstream to downstream through the wavefront, the velocity decreases and the density and temperature increase. There exists a conversion between kinetic energy and internal energy, which is the effect of shock wave compression. At the same time, the electric field force does work on the ions, preventing the ions from moving downstream, so that part of the kinetic energy of the ions is converted into electric potential energy, which is the effect of the electric field force. For ion density, these two mechanisms (electric field force and shock wave compression) compete with each other, making the ion density peak appear an inflection point. When the Ma number is less than the inflection point Ma, as the Ma number increases, the enhancement of the electric field force dominates, causing the density peak to increase; when the Ma number is higher than the inflection point Ma, the enhancement of shock wave compression dominates, making the density peak decrease. For temperature peaks and velocity peaks, it is also necessary to consider the energy conversion between kinetic energy, internal energy and electric potential energy. Therefore, the structure of density peak shows a different tendency from other quantities.

Figure \ref{Fig:Ele3} shows the electrical quantity peak values versus $\rm{Ma}$. It can be seen from Figure \ref{Fig:Ele3} (a) that the electric field force remains growing as $\rm{Ma}$ increases, indicating that the extent of charge separation increases. The electric field force is the expression of the non-uniform charge distribution inside the shock wave, so there exists strong charge non-uniform phenomenon inside the shock wave, which greatly affects the shock wave structure. From Figures (b) and (c), it is observed that the absolute values of the positive and negative peak keep growing, but the growth rates of both keep decreasing gradually.

\subsection{Non-equilibrium Effects around plasma shock wave}
\begin{figure}
\centering
\includegraphics[height=7cm]{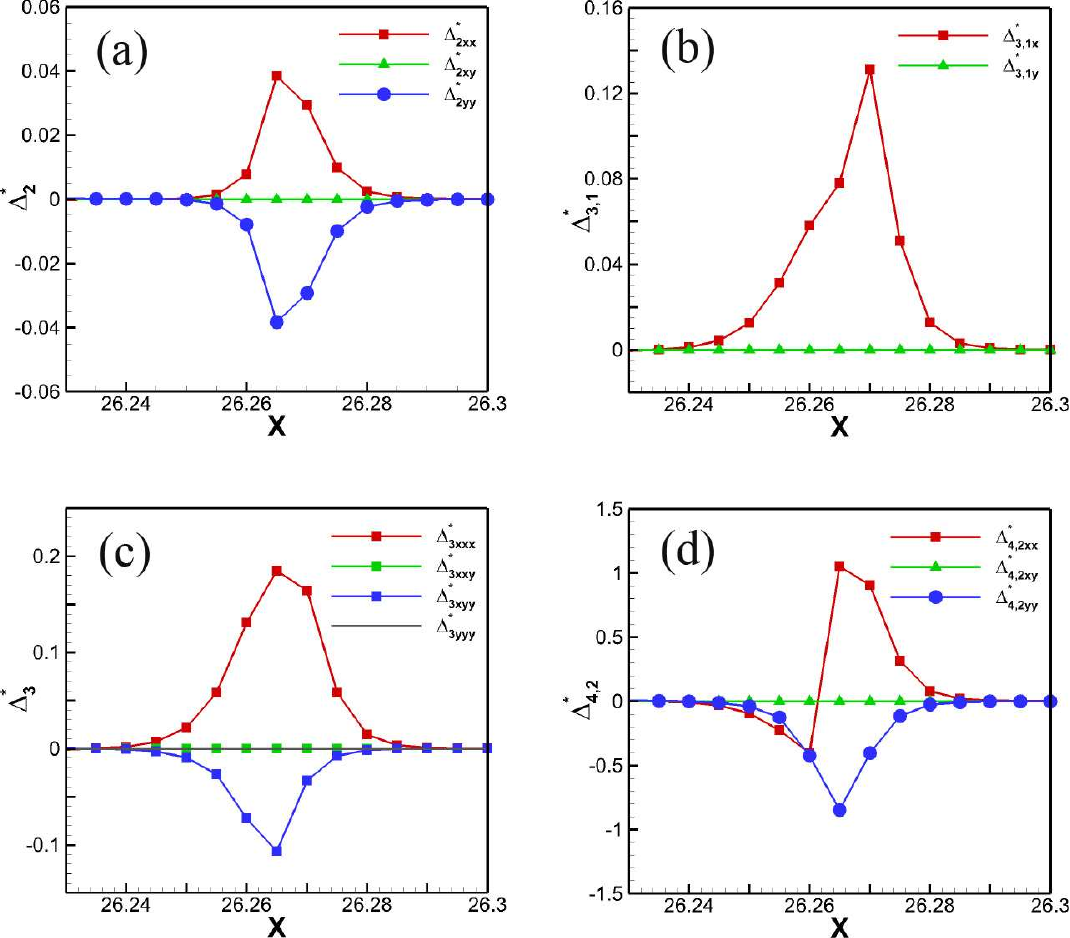}
\caption{Non-equilibrium quantities versus $x$ when $\rm{Ma}=1.8$. (a) $\Delta_2^*$, (b) $\Delta_{3,1}^*$, (c) $\Delta_{3}^*$, (d) $\Delta_{4,2}^*$.}
\label{Fig:Ma1.8TNEPlasma}
\end{figure}
Figure \ref{Fig:Ma1.8TNEPlasma} shows the non-equilibrium quantities of ion when Ma=1.8. Among the different types of TNE defined by Eq. \eqref{Eq:TNE2}, the most commonly used are non-organized momentum flux (NOMF) $\boldsymbol{\Delta}_2^*$ and non-organized energy flux (NOEF) $\boldsymbol{\Delta}_{3,1}^*$. The former have three components including $\Delta_{2xx}^*$, $\Delta_{2xy}^*$ and $\Delta_{2yy}^*$. $\Delta_{2xx}^*$ and $\Delta_{2yy}^*$ indicate the momentum flux in $x$ and $y$ direction, respectively. While $\Delta_{2xy}^*$ indicates the shear effect. $\boldsymbol{\Delta}_{3,1}^*$ represents the energy flux, and its two components $\Delta_{3,1x}^*$ and $\Delta_{3,1y}^*$ indicate the energy flux in $x$ and $y$ direction, respectively. From Figure \ref{Fig:Ma1.8TNEPlasma} (a) it could be found that the NOMF was symmetrically distributed in the $x$ and $y$ directions, which means the way system deviate from equilibrium in $x$ and $y$ direction is similar but towards different direction. It should be note that, the way system deviate from equilibrium in three-dimensional case can also be inferred from two-dimensional results. Due to the symmetry of the system, we cannot assume that the way the system deviates from equilibrium in the $y$ and $z$ directions are different, so the $\Delta_{2xx}^*$ is the same for two and three dimensional case. However, $\Delta_{2yy}^*$ will be evenly distributed in the $y$ and $z$ directions for three-dimensional case, and the sum of these three components is zero. Besides, $\Delta_{2xy}^*$ is zero, indicating that there is no shear effect. Figure \ref{Fig:Ma1.8TNEPlasma} (c) gives the distribution of $\boldsymbol{\Delta}_{3,1}^*$. It is observed that $\Delta_{3,1x}^*$ always deviate from equilibrium in one direction, and reaches its maximum at the wave front. $\Delta_{3,1y}^*$ is zero, which means there exist no non-equilibrium effect of energy flux in $y$ direction. $\boldsymbol{\Delta}_{3}^*$ and $\boldsymbol\Delta_{4,2}^*$ are correspond to flux of viscous effect and heat flux. From Figure \ref{Fig:Ma1.8TNEPlasma} (b), it is observed that the flux of $\Delta_{2xx}^*$ and $\Delta_{2yy}^*$ in $x$ direction is not zero, and both deviate from equilibrium toward one direction. However, the magnitude of $\Delta_{3xxx}^*$ is larger than $\Delta_{3xyy}$. Figure \ref{Fig:Ma1.8TNEPlasma} (d) describes the flux of $\boldsymbol{\Delta}_{3,1}^*$. It is found that the flux of $\Delta_{3,1x}^*$ in $x$ direction appears a reverse at downstream, but the flux of $\Delta_{3,1y}^*$ in $y$ direction always toward one direction.

\begin{figure}
\centering
\includegraphics[height=7cm]{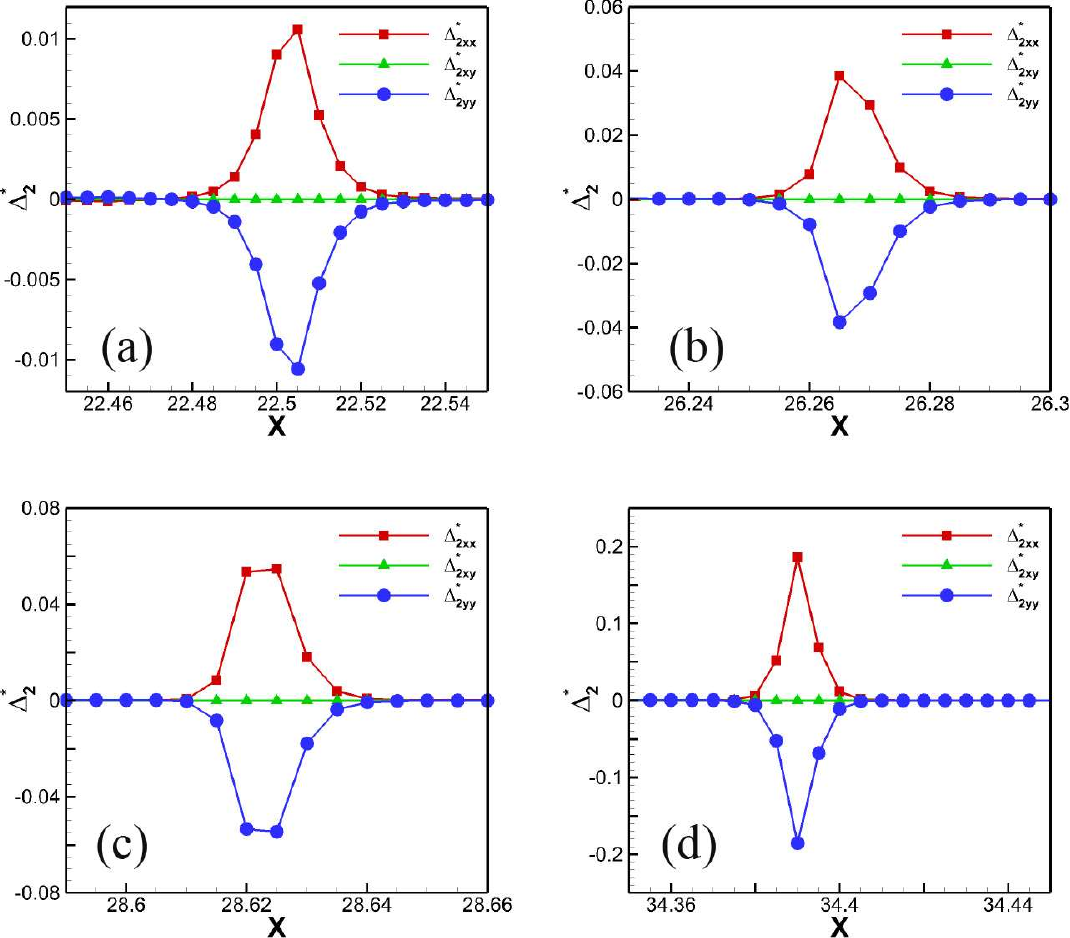}
\caption{\bm{$\Delta_2^*$} varies with $\rm{Ma}$. (a) Ma=1.5, (b) Ma=1.8, (c) Ma=2.0, (d) Ma=2.5.}
\label{Fig:Delta2}
\end{figure}
\begin{figure}
\centering
\includegraphics[height=7cm]{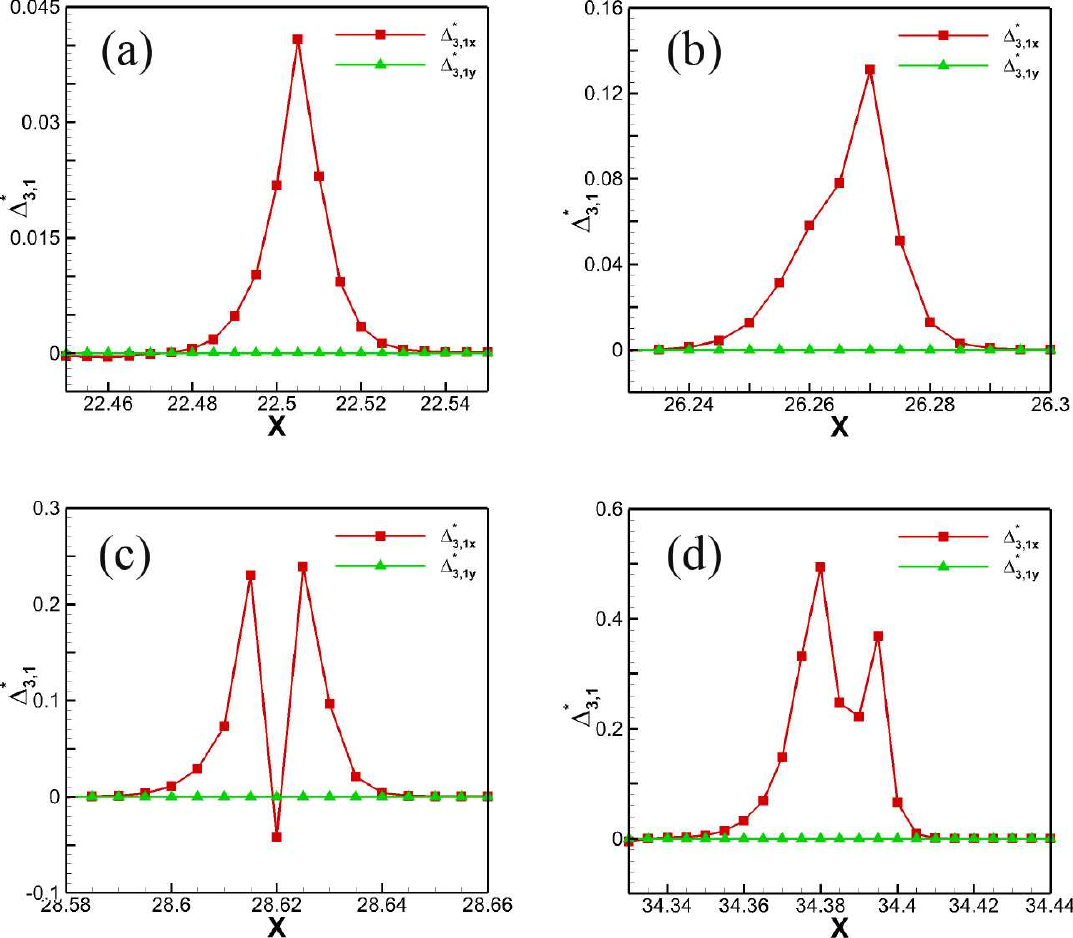}
\caption{\bm{$\Delta_{3,1}^*$} varies with $\rm{Ma}$. (a) Ma=1.5, (b) Ma=1.8, (c) Ma=2.0, (d) Ma=2.5.}
\label{Fig:Delta31}
\end{figure}
\begin{figure}
\centering
\includegraphics[height=7cm]{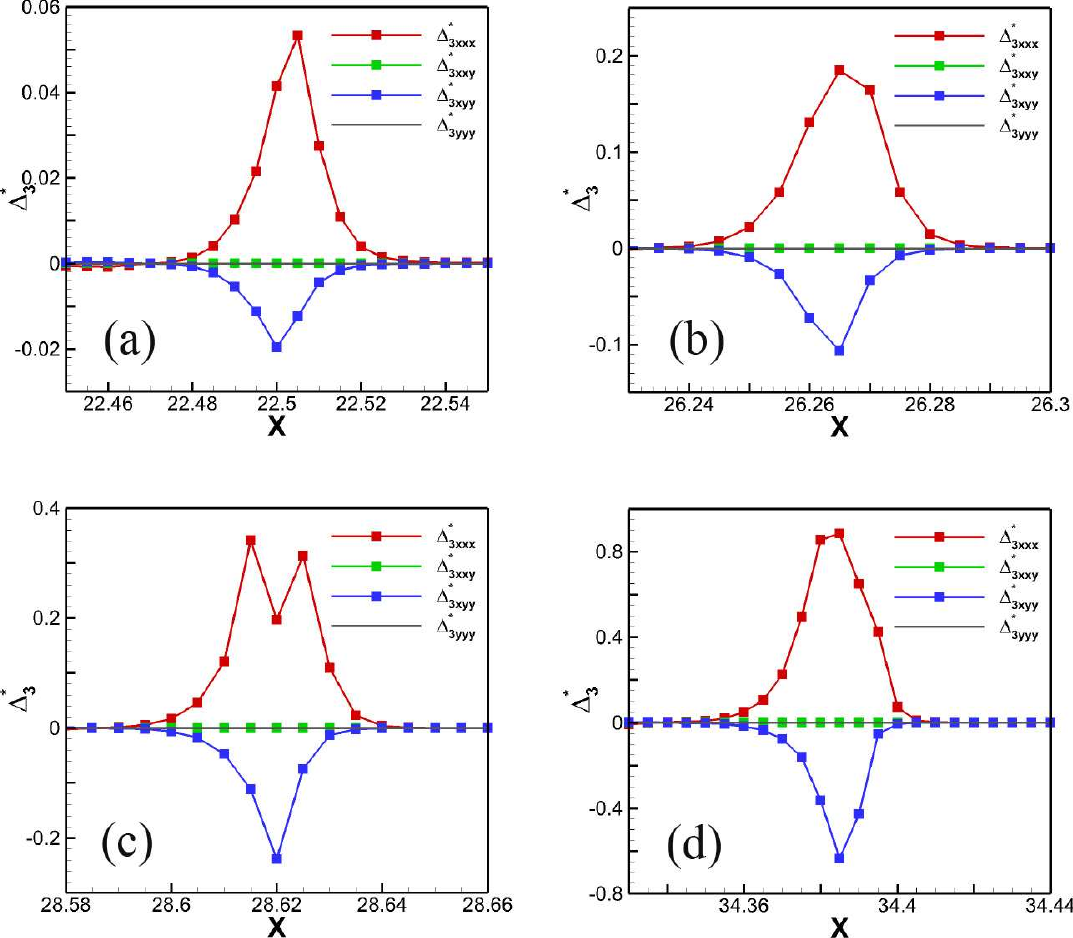}
\caption{\bm{$\Delta_3^*$} varies with $\rm{Ma}$. (a) Ma=1.5, (b) Ma=1.8, (c) Ma=2.0, (d) Ma=2.5.}
\label{Fig:Delta3}
\end{figure}
\begin{figure}
\centering
\includegraphics[height=7cm]{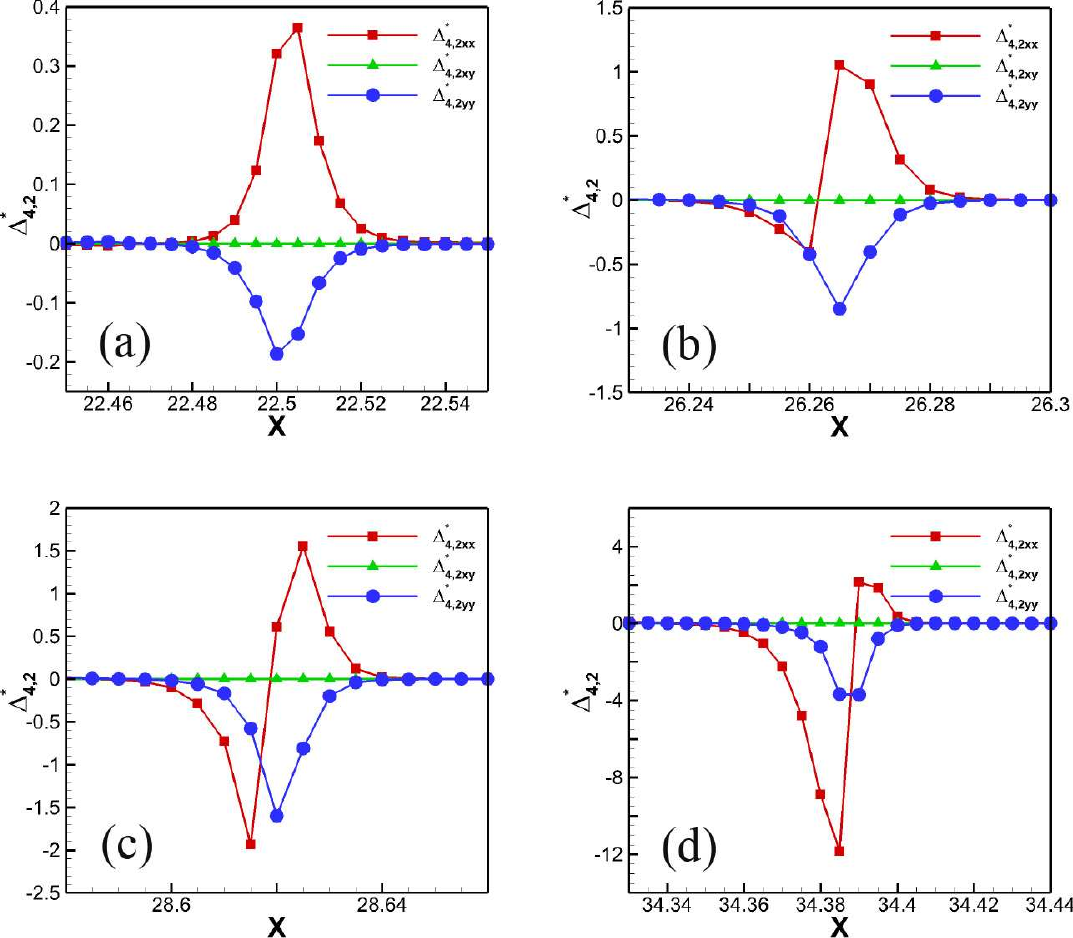}
\caption{\bm{$\Delta_{4,2}^*$} varies with $\rm{Ma}$. (a) Ma=1.5, (b) Ma=1.8, (c) Ma=2.0, (d) Ma=2.5.}
\label{Fig:Delta42}
\end{figure}

By changing the magnitude of the Mach number, we further investigate the variation of the non-equilibrium effects with Mach number. Some simulation results are shown in Figures \ref{Fig:Delta2} - \ref{Fig:Delta42} which are for $\boldsymbol{\Delta}^*_2$, $\boldsymbol{\Delta}^*_{3,1}$, $\boldsymbol{\Delta}^*_3$ and $\boldsymbol{\Delta}^*_{4,2}$, respectively. Firstly, the amplitudes of all the four quantities increase with $\rm{Ma}$. Then, from Figure \ref{Fig:Delta31}, it is observed that the NOEF in $x$ direction $\boldsymbol{\Delta}^*_{3,1x}$ appears a reverse when $\rm{Ma}=2.0$. When $\rm{Ma}=2.5$, $\boldsymbol{\Delta}^*_{3,1x}$ become positive again, forming a bimodal structure, which means the increasing $\rm{Ma}$ causes the strong non-equilibrium of energy flux in $x$ direction.
Finally, an evident difference from the case of detonation wave is that $\boldsymbol{\Delta}^*_{4,2}$ is qualitatively different from $\boldsymbol{\Delta}^*_2$ for the plasma shock wave\cite{Yan2013FOP}.

The difference between detonation wave and ordinary shock wave is that the former contains the contribution of chemical reaction. Therefore, the understanding of detonation wave structure can start from the ordinary shock wave structure. The density, velocity, temperature and pressure before and after ordinary shock wave are determined by the Rankine-Hugoniot relation. The relation describes the conservations of mass, momentum and energy between wave front and wave back. From the physical quantity distribution morphology, compared with ordinary fluid shock wave, detonation wave has a von Neumann peak at the wave-front and wave-back interface. The densities, flow velocities, temperatures and pressures of on the platforms of both sides, where the system is in equilibrium state, satisfy the Rankine-Hugoniot relation of detonation wave. As shown in Figures 10-14 in Reference\cite{Yan2013FOP}. Let's understand the reasons for the formation of von Neumann peak:

In the case of detonation wave, due to the compression of shock wave, the density, velocity, temperature and pressure will increase rapidly and reach their von Neumann peak values. In this process, chemical reaction begins to occur as soon as the temperature is raised to a threshold temperature for the chemical reaction to start. The leading and trailing edges of the von Neumann peak cross the reaction region, that is, the region where the reaction heat is released. The mode of exothermic heat is described by the reaction rate equation. In that case\cite{Yan2013FOP}, the rate of exothermic heat first increased and then decreased with time/space. When the chemical reaction is over, no more heat is released. The system state relaxes rapidly to the stationary equilibrium state behind the von Neumann peak.

In the case of reaction heat release, the pressure naturally becomes higher. The pressure in the reaction zone produces pressure difference relative to both sides of the reaction zone. The pressure gradient forces on both sides produce acceleration. The pressure gradient relative to the wave front is larger, so the forward acceleration is larger. (1) The accelerations in opposite directions on both sides of the reaction zone produce compression effects. This is why the temperature, density and pressure of the reaction zone are higher than those of the two sides. The forward acceleration is greater, making the von Neumann front steeper. (2) The interpretation that the flow velocity in the reaction zone is larger than that in the stationary region behind the wave is as follows. We roughly regard the reaction zone as a rigid body structure for the moment, and the acceleration forward of the front edge of the reaction zone is greater than that behind the stationary region behind the wave, so the net acceleration is forward.

The fine physical structure of the detonation wave is described by a non-conservative kinetic moments of $(f-f^{eq})$. As shown in Figures. 10-14 of reference\cite{Yan2013FOP}, $\boldsymbol\Delta_2^*$ and $\boldsymbol\Delta_{4,2}^*$ have different dimensions and physical meanings, but their spatial distribution forms are similar, that is, these corresponding kinetic properties deviated from thermodynamic equilibrium in similar ways.

As shown in Figures \ref{Fig:Ma1.5}(a) and \ref{Fig:Ma2.0}(a) of the current manuscript, in the structure of plasma shock wave, the distributions of density, flow rate, temperature and pressure show similar von Neumann peaks, but the spatial distributions of $\boldsymbol\Delta_2^*$ and $\boldsymbol\Delta_{4,2}^*$ are obviously different, as shown in Figure \ref{Fig:Ma1.8TNEPlasma} of the current manuscript.

The mechanisms of forming peak structure in the two cases are completely different. In the case of detonation wave, the peak results from exothermic heat by chemical reaction, while in the case of plasma shock wave, the peak results are due to the action of electric field force.

To be specific, we investigated the HNE and TNE characteristics of detonation wave by using DBM and the Lee-Tarver chemical reaction model\cite{Yan2013FOP}. It is found that the macroscopic quantities both show spike structures, and the $\boldsymbol{\Delta}^*_{2}$ and $\boldsymbol{\Delta}^*_{4,2}$ exhibit similar characteristics where the $xx$ and $yy$ components of these two quantities deviate from its equilibrium in opposite directions with the same deviation amplitude before and after shock wave front. However, it is obvious from Figure \ref{Fig:Delta2} and Figure \ref{Fig:Delta42} that these two quantities show different characteristics.

\section{Conclusion}
A discrete Boltzmann model for plasma shock wave is constructed. The model works for both steady and un-steady shock waves.
The electron is assumed inertialess and always in thermodynamic equilibrium. The Rankine-Hugoniot relations for single fluid theory of plasma shock wave is derived. It is found that the physical structure of shock wave in plasma is significantly different from that in neutral fluid and somewhat similar to that of detonation wave from the sense that a peak appears in the front. The charge of electricity deviates oppositely from neutrality in upstream and downstream of the shock wave. The large inertia of the ions causes them to lag behind, so the wave front charge is negative and the wave rear charge is positive.
The non-equilibrium effects around the shock front become stronger with increasing Mach number.
The variations of HNE and TNE with Mach number are numerically investigated. The characteristics of TNE can be used to distinguish plasma shock wave from detonation wave.

 It is understandable that the dissipative effects of electron and the diffusion effects between ion and electron have not been taken into account in this work. It is still an open topic that DBM for more practical cases where both the electron thermal conductivity and momentum/energy transfer between ion and electron are important. The two fluid DBM for plasma is in progress, which will be published in future.

Finally, it should be noted that, like the NS model, DBM is a theoretical model to describe the behavior of the coarse-grained system, and the scope and depth of its description of the dynamic properties of the system can be adjusted according to the research needs of specific problems.The DBM is the same as the NS model, and it is necessary to select the appropriate numerical calculation method before the numerical experiment is carried out. From the point of view of the research needs of physical problems, we can choose any numerical calculation method that meets the research needs of physical problems and is allowed by the current hardware and software environment. Readers interested in numerical methods may refer to more specialized literature.

\begin{acks}

The authors thank Lifeng Wang, Hongbo Cai, Yingkui Zhao for helpful discussions on plasma and implosion physics, thank Yanbiao Gan, Feng Chen, Chuandong Lin, Ge Zhang, Guanglan Sun, Jie Chen, Dejia Zhang, Yiming Shan, Hanwei Li, and Cheng Chen for helpful discussions on DBM, thank Long Miao, Song Bai, Dongfeng Yan, Xiaolong Yi and Fuwen Liang for helpful discussions on the article organization.
This work was supported by the National Natural Science Foundation of China (under Grant Nos. 12172061, 12102397, and 11602162),
CAEP Foundation (under Grant No. CX2019033), the opening project of State Key
Laboratory of Explosion Science and Technology (Beijing Institute of Technology) (under Grant No. KFJJ21-16M), and the China Postdoctoral Science Foundation (under Grant No. 2019M662521).
\end{acks}

\end{document}